\documentclass[
superscriptaddress,
preprint,
prd,tightenlines,showpacs,nofootinbib,
eqsecnum,
amsfonts,amsmath,amssymb]{revtex4-1}

\usepackage{bm}
\usepackage{graphicx} 
\usepackage{hyperref}
\usepackage{mathrsfs}
\usepackage{multirow}

\newcommand{\bfy}{\mathbf{y}}
\newcommand{\bfx}{\mathbf{x}}
\newcommand{\bfV}{\mathbf{V}}
\newcommand{\bfv}{\mathbf{v}}

\newcommand{\bfp}{\mathbf{p}}
\newcommand{\bfS}{\mathbf{S}}
\newcommand{\ov}[1]{\overline{#1}}
\newcommand{\ud}{\mathrm{d}}
\newcommand{\uD}{\mathrm{D}}
\newcommand{\ui}{\mathrm{i}}

\newcommand{\calO}{\mathcal{O}}
\newcommand{\calF}{\mathcal{F}}
\newcommand{\ph}[1]{\phantom{#1}}

\usepackage{ulem}
\usepackage[usenames]{color}

\allowdisplaybreaks

\begin{document}

\title{Next-to-next-to-leading order spin-orbit effects in the equations of
  motion of compact binary systems}

\author{Sylvain \textsc{Marsat}}\email{marsat@iap.fr}
\affiliation{$\mathcal{G}\mathbb{R}\varepsilon{\mathbb{C}}\mathcal{O}$
  Institut d'Astrophysique de Paris --- UMR 7095 du CNRS, \\ Universit\'e
  Pierre \& Marie Curie, 98\textsuperscript{bis} boulevard Arago, 75014 Paris,
  France}

\author{Alejandro \textsc{Boh\'{e}}}\email{bohe@iap.fr}
\affiliation{$\mathcal{G}\mathbb{R}\varepsilon{\mathbb{C}}\mathcal{O}$
  Institut d'Astrophysique de Paris --- UMR 7095 du CNRS, \\ Universit\'e
  Pierre \& Marie Curie, 98\textsuperscript{bis} boulevard Arago, 75014 Paris,
  France}
\affiliation{Departament de F\'isica, Universitat de les Illes Balears, Crta.
  Valldemossa km 7.5, E-07122 Palma, Spain}

\author{Guillaume \textsc{Faye}}\email{faye@iap.fr}
\affiliation{$\mathcal{G}\mathbb{R}\varepsilon{\mathbb{C}}\mathcal{O}$
  Institut d'Astrophysique de Paris --- UMR 7095 du CNRS, \\ Universit\'e
  Pierre \& Marie Curie, 98\textsuperscript{bis} boulevard Arago, 75014 Paris,
  France}

\author{Luc \textsc{Blanchet}}\email{blanchet@iap.fr}
\affiliation{$\mathcal{G}\mathbb{R}\varepsilon{\mathbb{C}}\mathcal{O}$
  Institut d'Astrophysique de Paris --- UMR 7095 du CNRS, \\ Universit\'e
  Pierre \& Marie Curie, 98\textsuperscript{bis} boulevard Arago, 75014 Paris,
  France}

\date{\today}

\begin{abstract}
  We compute next-to-next-to-leading order spin contributions to the
  post-Newtonian equations of motion for binaries of compact objects, such as
  black holes or neutron stars. For maximally spinning black holes, those
  contributions are of third-and-a-half post-Newtonian (3.5PN) order,
  improving our knowledge of the equations of motion, already known for
  non-spinning objects up to this order. Building on previous work, we
  represent the rotation of the two bodies using a pole-dipole matter
  stress-energy tensor, and iterate Einstein's field equations for a set of
  potentials parametrizing the metric in harmonic coordinates. Checks of the
  result include the existence of a conserved energy, the approximate global
  Lorentz invariance of the equations of motion in harmonic coordinates, and
  the recovery of the motion of a spinning object on a Kerr background in the
  test-mass limit. We verified the existence of a contact transformation,
  together with a redefinition of the spin variables that makes our result
  equivalent to a previously published reduced Hamiltonian, obtained from the
  Arnowitt-Deser-Misner (ADM) formalism.
\end{abstract}


\maketitle


\section{Introduction}
\label{Introduction}

Gravitational wave detectors will soon enter a new era, with the advanced
versions of Virgo and LIGO expected to start operating around 2015, and the
construction of KAGRA \cite{virgo, ligo, KAGRA}. Primary targets for these
laser ground-based interferometers, or for possible future space-based
interferometers, are inspiralling binaries of compact objects --- neutron
stars and/or black holes ---, which emit gravitational radiation during their
late inspiral, merger, and ringdown phases. Hunting for the faint signal and
best separating it from the noise using matched-filtering techniques
(\textit{e.g.} \cite{Cutler&al93, CF94})
requires a very precise modelling of the expected waveform, which the
post-Newtonian  (hereafter PN) approximation aims at providing
for the inspiral phase \cite{Bliving}.

Up to now, such high-precision PN templates are available for the non-spinning
case up to the 3.5PN order, \textit{i.e.} the order $1/c^7$ in the formal
expansion in powers of $1/c^2$ (with $c$ being the speed of light). However,
observational evidence points toward the existence of fast-rotating black
holes, endowed with significant angular momentum (or spin), for stellar size
\cite{AK01, Strohmayer01, 2006ApJ...652..518M, Gou&al11, 2012AIPC.1427...48N}
as well as for supermassive black holes \cite{FM05, BR06, Brenneman+11}.

It is thus important to complete our knowledge of the PN predictions for the
gravitational waves emitted by such systems by including the spins of each of
the two bodies. This implies first taking those spins into account in the
dynamics of the binary. For maximally spinning objects, the leading-order
linear-in-spin (which we shall call spin-orbit) effect arises at 1.5PN order
when regarding the spin as a 0.5PN quantity [see Eq.~\eqref{spinPNcounting}
below]. The next-to-leading and next-to-next-to-leading contributions show up
at respectively 2.5PN and 3.5PN orders. The effect of spacetime curvature on
the motion of a spinning test particle was obtained in the seminal work by
Papapetrou \cite{Papa51,Papa51spin,CPapa51spin}, after an earlier derivation
by Mathisson (republished in \cite{Mathisson37repub}). 
Barker and O'Connell \cite{BOC75, BOC79} determined the
leading-order spin-orbit and spin-spin effects in the two-body dynamics. More
recently, Kidder, Will and Wiseman \cite{KWW93, Kidder95} computed the
corresponding contributions to the radiation field and, most importantly, to
the orbital phase of the binary, to which the templates are crucially
sensitive. Effective field theory methods \cite{GR06} were also used to
rederive the leading-order spin-orbit and spin-spin contributions to the
dynamics \cite{Porto06}. The problem in the limit of a spinning
test particle on a Kerr background has been addressed in \cite{MST96, TMSS96}
(see also \cite{BRB09} for a Hamiltonian model neglecting the gravitation
damping force).

By means of a PN iteration of the Einstein field equations in harmonic
coordinates, the equation of motion at the next-to-leading order was first
investigated by Tagoshi, Ohashi and Owen \cite{OTO98, TOO01}. It was then
confirmed and completed (as well as extended to the radiation field) by
Blanchet, Buonanno and Faye \cite{FBB06, BBF06}. The results for the evolution
equations were retrieved by two independent calculations, using a Hamiltonian
approach in ADM coordinates \cite{DJSspin} on the one hand, and using
effective field theory methods \cite{Levi10, Porto10} on the other hand. The
ADM computation was later generalized to the many-body problem \cite{HS11} and
extended to the next-to-leading order spin1-spin2 and spin square
(\textit{e.g.} spin1-spin1) effects in Refs.~\cite{SHS07b, HSS10}.
Next-to-next-to-leading order spin-orbit effects in the Hamiltonian of the
binary were first computed in \cite{HS11spinorbit},
resorting to the ADM scheme adapted to matter sources composed of spinning
point particles \cite{SS09a}. In a subsequent work, spin1-spin2 interactions
terms were also added at the 4PN order \cite{HS11spinspin}. Effective
field theory methods progressed concurrently by computing the 3PN spin1-spin2
and spin1-spin1 contributions \cite{PR08a, PR08b, Levi08}, and the 4PN
spin1-spin2 interactions \cite{Levi12}.

Our aim in this work is to extend the approach of
Ref.~\cite{FBB06},\footnote{We will refer to Ref.~\cite{FBB06} as Paper~I
  throughout this work.} based on the PN expansion of the metric in harmonic
coordinates, in order to compute the spin-orbit next-to-next-to-leading order
3.5PN contributions to the equations of motion. We shall thus provide an
essential validation of the result obtained by the ADM method
\cite{HS11spinorbit}. Our final purpose is to derive the spin-orbit terms in
the emitted GW energy flux and in the phase of the binary at the same
approximation, extending therefore Ref.~\cite{BBF06} to
next-to-next-to-leading order, so as to provide more accurate PN predictions
to the template-based data analysis.

We adopt the same convention for the PN order counting
of the spins as in Paper~I; namely, we redefine our spin variable with respect
to the ``true'' spin angular momentum following~:
\begin{equation}\label{spinPNcounting}
	S \equiv c S_{\text{true}}  = G m_\text{body}^{2} \chi \;,
\end{equation}
where $m_\text{body}$ is the mass and $\chi$ the dimensionless spin parameter
($G$ denotes Newton's constant). For a maximally rotating compact object, we
have $\chi \sim 1$ and $S \sim G m_\text{body}^{2}$, so that our spin variable
can be counted as of Newtonian order in this case. For slowly rotating objects
however, we have $\chi \sim v_\text{surf}/c$ where $v_\text{surf}$ is the
rotation velocity of the body surface, which implies that $S$ acquires an
additional factor $1/c$.
In that case, spin-orbit contributions effectively appear at 2PN order, and
spin-spin ones at 3PN only. Our counting will always assume rapid rotation.
The leading-order spin-orbit terms will thus carry an explicit factor
$1/c^3$; in the present article we shall obtain the next-to-leading and
next-to-next-to-leading spin-orbit corrections $\sim 1/c^5$ and $\sim 1/c^7$.

The paper is organized as follows. In Sec.~\ref{secPoledipole}, we review the
pole-dipole formalism for modelling compact objects as point particles with
spins. In Sec.~\ref{secPNmetric}, we describe the parametrization of the PN
metric in terms of a set of elementary potentials and present the equations of
motion as deduced generically from metric components. The core of our work
consists of the computation of the required potentials in
Secs.~\ref{secHadamard} and \ref{secDimreg}, first using Hadamard's
regularization (as well as the ``pure Hadamard-Schwartz'' prescription
\cite{BDE04}), and next completing some of our calculations by means of the
more powerful dimensional regularization. Sec.~\ref{secResults} contains our
results for the precession equations and the acceleration. In
Sec.~\ref{secChecks}, we explain the various checks that have been performed
to validate them: construction of a conserved energy, check of the global
Lorentz invariance, test-mass limit, and recovery of the results of
Ref.~\cite{HS11spinorbit} in the ADM Hamiltonian formulation. We conclude
shortly in Sec.~\ref{Conclusion}.


\section{Effective pole-dipole formalism}
\label{secPoledipole}

The starting point of our calculations is the model of pole-dipole particles
developed by Mathisson \cite{Mathisson37repub}, Papapetrou \cite{Papa51spin,
  CPapa51spin}, Tulczyjew \cite{Tulczyjew57, Tulczyjew59} and generalized by
Dixon \cite{Dixon64, Dixon73, Dixon79} and Bailey \& Israel \cite{BI80}. This
model allows an effective description of the dynamics of bodies that
accounts for their spin angular momentum by means of singular Dirac
delta-functions, making analytical computations tractable. We use the same
version of the model as in Paper~I, staying linear in the spins, but we
summarize here the main formulae for completeness and refer to Paper~I for
more details.

The central assumption is that each particle is described by a stress-energy
tensor made of two parts, a monopolar one and a dipolar one:
$T^{\mu\nu}=T^{\mu\nu}_\text{M}+T^{\mu\nu}_\text{D}$. These are built with
respectively a Dirac delta function and a gradient of a delta function,
integrated over the world line of the particle, according to:
\begin{subequations}\label{Tmunu}
\begin{align}
	T^{\mu\nu}_\text{M} &= c^{2} \int^{+\infty}_{-\infty} \ud\tau \; 
p^{(\mu}u^{\nu)}\frac{\delta^{(4)}(x-y(\tau))}{\sqrt{-g(x)}} \;, \\
	T^{\mu\nu}_\text{D} &= -c \int^{+\infty}_{-\infty} \ud\tau \;
    \nabla_{\rho} \left[ S^{\rho(\mu}u^{\nu)}
\frac{\delta^{(4)}(x-y(\tau))}{\sqrt{-g(x)}} \right] \;.
\end{align}
\end{subequations}
Here $\tau$ is the proper time measured along the world line, described itself
by the particle position $y^\mu(\tau)$; $\delta^{(4)}$ denotes the
four-dimensional Dirac delta function, $g$ stands for the determinant of the
spacetime
metric $g_{\mu\nu}$; $u^{\mu}=\ud y^{\mu}/(c\ud\tau)$ is the four-velocity of
the particle (satisfying $u_\mu u^\mu=-1$), $p^{\mu}$ its linear momentum, and
$S^{\mu\nu}$ an antisymmetric tensor that represents the spin of the particle.

The linear momentum $p^{\mu}$ and spin tensor $S^{\mu\nu}$ obey the
Mathisson-Papapetrou equations of evolution:
\begin{subequations} \label{papapetrou}
\begin{align}
	\frac{\uD S^{\mu\nu}}{\ud\tau} &=
    c^{2}\bigl(p^{\mu}u^{\nu}-p^{\nu}u^{\mu}\bigr) 
\;,\label{papprecession} \\
	\frac{\uD p^{\mu}}{\ud\tau} &= -\frac{1}{2}
    R^{\mu}_{\ph{\mu}\nu\rho\sigma} u^{\nu}S^{\rho\sigma}\;, \label{papeom}
\end{align}
\end{subequations}
with $\uD/\ud\tau\equiv cu^{\nu}\nabla_{\nu}$, and
$R^{\mu}_{\ph{\mu}\nu\rho\sigma}$ denoting the Riemann tensor. As is well
known, a choice of supplementary spin condition (thereafter SSC) is necessary,
in order to obtain the correct number of degrees of freedom (see \cite{KS07}
for a summary of the various choices in use in the litterature). We
adopt the following covariant SSC:
\begin{equation}\label{SSC}
	S^{\mu\nu}p_{\nu} = 0\;.
\end{equation}
With this relation in hand, one can combine the equations \eqref{papapetrou}
to obtain the link between the four-velocity $u^{\mu}$ and the linear
momentum $p^{\mu}$; hence we deduce the conservation laws:
\begin{equation}
	\frac{\uD m}{\ud\tau}=0 \;, \qquad \frac{\uD S}{\ud\tau}=0\;,
\end{equation}
for the mass and for the magnitude of the spin defined by
$m^{2}c^{2}=-p^{\mu}p_{\mu}$ and $S^{2}=S^{\mu\nu}S_{\mu\nu}/2$ respectively,
and thus conserved along the particle's world line.

We now restrict the evolution equations themselves to linear order in spins,
neglecting any term $\calO(S^{2})$, quadratic or of higher order. One can
readily check the proportionality relation between $u^{\mu}$ and $p^{\mu}$ in
that approximation:
\begin{equation}\label{relationup}
	p^{\mu}=m\,c \,u^{\mu} + \calO(S^{2}) \;.
\end{equation}
In particular, the SSC now reads $S^{\mu\nu}u_{\nu}=\calO(S^{3})$. Notice that
the monopolar part of the stress-energy tensor has the same form as the one of
an ordinary point-like particle, but it does depend on the spin implicitly through the metric. Equations \eqref{papapetrou} become:
\begin{subequations} \label{papapetroulinear}
\begin{align}
	\frac{\uD S^{\mu\nu}}{\ud\tau} &=
    \calO(S^{2}) \label{papprecessionlinear} \;, \\
	mc\frac{\uD u^{\mu}}{\ud \tau} &= -\frac{1}{2}
    R^{\mu}_{\ph{\mu}\nu\rho\sigma}u^{\nu}S^{\rho\sigma} +\calO(S^{2})
    \;, \label{papeomlinear}
\end{align}
\end{subequations}
We see that in the linear approximation, the spin tensor is parallely
transported along the motion, which is actually non-geodesic due to the
coupling of the spin tensor with the Riemannian curvature.

In the following, we shall use the 3-dimensional form of the energy-momentum
tensor \eqref{Tmunu}. Using a $3+1$ splitting of spacetime, the particle's
position and coordinate velocity are denoted $y^\mu=(c\,t, \mathbf{y}(t))$ and
$v^\mu(t)=(c, \mathbf{v}(t))$ (where $v^{\mu}=cu^{\mu}/u^{0}$, with
$u^{0}=1/\sqrt{-g_{\rho\sigma}v^{\rho}v^{\sigma}/c^{2}}$), and the spin tensor
$S^{\mu\nu}(t)$ is considered a function of time. From now on, we use
boldface letters to denote three-dimensional vectors. We have
\begin{subequations}\label{Tmunu3D}
\begin{align}
	T^{\mu\nu}_\text{M} &= m u^{0}v^{\mu}v^{\nu}
\frac{\delta^{(3)}(\mathbf{x}-\mathbf{y}(t))}{\sqrt{-g(t,\mathbf{x})}}\;,\\
	T^{\mu\nu}_\text{D} &= -\frac{1}{c} \nabla_{\rho}
\left[ S^{\rho(\mu}v^{\nu)}
\frac{\delta^{(3)}(\mathbf{x}-\mathbf{y}(t))}{\sqrt{-g(t,\mathbf{x})}}
 \right]\;,
\end{align}
\end{subequations}
where $\delta^{(3)}$ is the three-dimensional Dirac delta function.
Expliciting the covariant derivative in the dipolar stress-energy tensor, we
obtain an alternative form in terms of ordinary Christoffel symbols:
\begin{equation}\label{Tmunudipolar}
	\sqrt{-g}\,T^{\mu\nu}_\text{D}= -\frac{1}{c} \left( \partial_{\rho} 
\Bigl[ S^{\rho(\mu}v^{\nu)}\delta^{(3)}(\mathbf{x}-\mathbf{y}(t)) \Bigr] + 
S^{\rho(\mu}\Gamma^{\nu)}_{\rho\sigma}v^{\sigma}\delta^{(3)}(\mathbf{x}-\mathbf{y}(t)) 
\right) \;.
\end{equation}
Note that in \eqref{Tmunu3D}--\eqref{Tmunudipolar} the factor $\sqrt{-g}$ has
the generic field point $(c\, t,\mathbf{x})$ for argument. Because of the SSC,
we may work only with the spatial components $S^{ij}$ of the spin tensor, and
eliminate the $S^{0i}$ components according to:
\begin{equation}\label{S0itoSij}
	u_{0}S^{0i} =-u_{j}S^{ji} + \calO(S^{3}) \;. 
\end{equation}

In Paper~I, a spin vector was used instead of a spin tensor. First, a spin
covector was defined, consistently with the SSC
$S^{\mu\nu}u_{\nu}=\calO(S^{3})$, by:
\begin{equation}\label{Smu}
	S^{\mu\nu} = \varepsilon^{\mu\nu\rho\sigma} u_{\rho} S_{\sigma} +  \calO(S^{3})\;, 
\end{equation}
where $\varepsilon^{\mu\nu\rho\sigma}$ is the Levi-Civita tensor such that
$\varepsilon^{0123}=-1/\sqrt{-g}$. Then, a spatial spin vector
$S^{i}_\text{FBB}$ was constructed as $S^{i}_\text{FBB}=\gamma^{ij}S_{j}$,
with $\gamma^{ij}$ being the inverse of the spatial part of the metric,
\textit{i.e.} $\gamma^{ik}g_{kj}=\delta^{i}_{\ph{i}j}$. In the present paper
we shall mostly work with the spatial components $S^{ij}$ of the spin tensor.
This presents two advantages: first, it somewhat simplifies the algebra and
the index structure by getting rid of the Levi-Civita tensors, and second, it
makes the check of the Lorentz invariance more straightforward, since $S^{ij}$
is directly the spatial components of the tensor $S^{\mu\nu}$. We provide in
Appendix \ref{appSiSij} the explicit link between $S^{ij}$ and the spin
variable $S^{i}_\text{FBB}$.


\section{Post-Newtonian metric and equations of motion}
\label{secPNmetric}

As in Paper~I, we use harmonic (or De Donder) coordinates, defined by the
gauge condition $\partial_{\nu}(\sqrt{-g}g^{\mu\nu})=0$. The PN iteration of
Einstein's field equations developed up to 3.5PN order in Ref.~\cite{BFeom} is
valid for a generic matter source, the only hypothesis being that the
stress-energy tensor $T^{\mu\nu}$ must have compact support. We will need here
the full 3.5PN metric, which is parametrized in terms of elementary potentials
as:\footnote{We denote by $\calO(n)$ remainder terms of order $(n/2)$-PN,
  \textit{i.e.} behaving as
$\mathcal{O}(1/c^{n})$.}
\allowdisplaybreaks{\begin{subequations}\label{metricg}
\begin{align} g_{00}
 &=  -1 + \frac{2}{c^{2}}V - \frac{2}{c^{4}} V^{2} + \frac{8}{c^{6}}
\left(\hat{X} + V_{i} V_{i} + \frac{V^{3}}{6}\right)\nonumber\\ & +
\frac{32}{c^{8}} \left(\hat{T} - \frac{1}{2} V \hat{X} + \hat{R}_{i} V_{i} -
\frac{1}{2} V V_{i} V_{i} - \frac{V^{4}}{48}\right)+\calO(10)\;,\\ 
g_{0i} & = - \frac{4}{c^{3}}
V_{i} - \frac{8}{c^{5}} \hat{R}_{i} - \frac{16}{c^{7}} \left(\hat{Y}_{i} +
\frac{1}{2}\hat{W}_{ij} V_{j} + \frac{1}{2} V^{2} V_{i}\right) + \calO(9)\;,\\ 
g_{ij} & = \delta_{ij} \left[1 +
\frac{2}{c^{2}}V + \frac{2}{c^{4}} V^{2} + \frac{8}{c^{6}} \left(\hat{X} +
V_{k} V_{k} + \frac{V^{3}}{6}\right)\right] \nonumber\\ & +
\frac{4}{c^{4}}\hat{W}_{ij} + \frac{16}{c^{6}} \left( \hat{Z}_{ij} + \frac{1}{2} V
\hat{W}_{ij} - V_{i} V_{j} \right) + \calO(8) \;.
\end{align}
\end{subequations}}\noindent
From now on, spatial indices $i,j,\dots$ will be raised and lowered using the
Kronecker metric $\delta_{ij}$, and we will write them indifferently as upper
or lower indices. The potentials are defined by means of the usual retarded
($\mathcal{R}$) inverse flat d'Alembertian operator,
\begin{equation}\label{defdalembertian}
	(\Box_{\mathcal{R}}^{-1}f)(\mathbf{x},t) = - \frac{1}{4\pi} \int 
\frac{\ud^{3}\mathbf{x}'}{|\mathbf{x}-\mathbf{x}'|}f
\left(\mathbf{x}',t-\frac{|\mathbf{x}-\mathbf{x}'|}{c}\right) \; ,
\end{equation}
where the sources $f$ are built with the following matter quantities,
\begin{equation}
	\sigma=\frac{1}{c^{2}}(T^{00}+T^{ii}) \; , \quad 
\sigma_{i}=\frac{1}{c}T^{0i} \; , \quad \sigma_{ij}=T^{ij} \; ,
\end{equation}
as well as with products of derivatives of lower-order potentials. The
potentials required for the 2.5PN equations of motion were already used in
Paper~I and read:
\begin{subequations}\label{defpotentials}
\allowdisplaybreaks{
\begin{align}
V & = \Box_{\mathcal{R}}^{-1}[-4 \pi G\, \sigma]\;,\label{V}\\ 
V_{i} &= \Box_{\mathcal{R}}^{-1}[-4 \pi G\, \sigma_{i}]\;,\\
\hat{X} &= \Box_{\mathcal{R}}^{-1}\left[\vphantom{\frac{1}{2}} -4 \pi G\, V 
\sigma_{ii} + \hat{W}_{ij}
\partial_{ij} V + 2 V_{i} \partial_{t} \partial_{i} V + V \partial_{t}^{2}
V\right.\nonumber\\ & \qquad\left. + \frac{3}{2}(\partial_{t} V)^{2} - 2
\partial_{i} V_{j}\partial_{j} V_{i}\right]\;,\\
\hat{R}_{i} & = \Box_{\mathcal{R}}^{-1}\left[-4 \pi G\, 
(V \sigma_{i} - V_{i} \sigma) - 2
\partial_{k} V \partial_{i} V_{k} - \frac{3}{2} \partial_{t} V \partial_{i}
V\right]\;, \\ 
\hat{W}_{ij} & =  \Box_{{\cal R}}^{-1}\left[-4 \pi G\,
(\sigma_{ij} - \delta_{ij} \sigma_{kk}) - \partial_{i} V \partial_{j}
V\right]\;,\label{Wij}
\end{align}
}
while the potentials required for the 3.5PN order are given the following
definitions:
%
\begin{align}
\hat{T} &= \Box_{\mathcal{R}}^{-1}\ \left[-4 \pi G\, \left(\frac{1}{4} \sigma_{ij}
\hat{W}_{ij} + \frac{1}{2} V^{2} \sigma_{ii} + \sigma V_{i} V_{i}\right) +
\hat{Z}_{ij} \partial_{ij} V + \hat{R}_{i} \partial_{t} \partial_{i} V \right.
\nonumber\\ 
&\qquad - 2 \partial_{i} V_{j} \partial_{j} \hat{R}_{i} -
\partial_{i} V_{j} \partial_{t} \hat{W}_{ij} + V V_{i} \partial_{t}
\partial_{i} V + 2 V_{i} \partial_{j} V_{i}\partial_{j} V + \frac{3}{2} V_{i}
\partial_{t} V \partial_{i} V \nonumber\\ 
&\qquad \left. + \frac{1}{2} V^{2}
\partial_{t}^{2} V + \frac{3}{2} V (\partial_{t} V)^{2} - \frac{1}{2}
(\partial_{t} V_{i})^{2} \right]\;,\\
\hat{Y}_{i} & = \Box_{{\cal R}}^{-1}\left[-4 \pi G\, 
\left(- \sigma \hat{R}_{i} - \sigma V V_{i} + \frac{1}{2}
\sigma_{k} \hat{W}_{ik} + \frac{1}{2} \sigma_{ik} V_{k} + \frac{1}{2}
\sigma_{kk} V_{i}\right) + \hat{W}_{kl} \partial_{kl} V_{i} \right.
\nonumber\\ 
&\qquad - \partial_{t} \hat{W}_{ik} \partial_{k} V +
\partial_{i} \hat{W}_{kl}\partial_{k} V_{l} - \partial_{k} \hat{W}_{il}
\partial_{l} V_{k} - 2 \partial_{k} V \partial_{i} \hat{R}_{k} - \frac{3}{2}
V_{k} \partial_{i} V \partial_{k} V \nonumber\\
&\qquad \left. - \frac{3}{2}
V \partial_{t} V \partial_{i} V - 2 V \partial_{k} V \partial_{k} V_{i} + V
\partial_{t}^{2} V_{i} + 2 V_{k} \partial_{k} \partial_{t} V_{i} \right]\;, \\
\hat{Z}_{ij} & = \Box_{\mathcal{R}}^{-1}\ \left[\vphantom{\frac{1}{2}}-4 \pi G\, V
\left(\sigma_{ij} - \delta_{ij} \sigma_{kk}\right) - 2 \partial_{(i} V
\partial_{t} V_{j)} + \partial_{i} V_{k} \partial_{j} V_{k} + \partial_{k}
V_{i}\partial_{k} V_{j} \right. \nonumber\\ 
&\qquad \left. - 2 \partial_{(i}
V_{k} \partial_{k} V_{j)} - \delta_{ij} \partial_{k} V_{m} (\partial_{k} V_{m}
- \partial_{m} V_{k}) - \frac{3}{4} \delta_{ij} (\partial_{t} V)^{2} \right]\;.
\end{align}
\end{subequations}
Notice the difference of structure between the sources of these potentials:
some sources are proportional to one of the $\sigma$ quantities and are
compact-supported, while others are only proportional to metric potentials. We
will call the latter non-compact supported, since their source extends in all
space.

Next, in keeping with notations of Paper~I, we rewrite the covariant equation
of motion \eqref{papeomlinear} in the following $3+1$ form:
\begin{equation}\label{eomstructure}
	\frac{\ud P_{i}}{\ud t} = F_{i} + \calF_{i} \; .
\end{equation}
The leading order of the left-hand side of the force law \eqref{eomstructure}
is simply the ordinary acceleration $a^{i}=\ud v^{i}/\ud t$. The
Newtonian-like linear momentum and forces are
\begin{subequations}\label{eomstructuredef}
\begin{align}
	P_{i} &= g_{i\nu} u^{0} v^{\nu}\;,\\
	F_{i} &= \frac{1}{2} \partial_{i} g_{\nu\rho} \,u^{0}v^{\nu}v^{\rho}\;,\\
	\calF_{i} &= - \frac{1}{2m c} R_{i\nu\rho\sigma}v^{\nu}S^{\rho\sigma}\; .
\label{eomrewrite2}
\end{align}\end{subequations}
With transparent meaning we shall often call $F_{i}$ the ``geodesic part'' of
the force law (or rather the acceleration), while $\calF_{i}$ will be referred
to as the ``Papapetrou part'', deviating from geodesic motion.

Expanding the metric in terms of the elementary potentials
\eqref{defpotentials}, one gets the following 3.5PN expressions for $P_{i}$
and $F_{i}$, which would correspond to the coordinate acceleration of a
particle following a geodesic motion:
\allowdisplaybreaks{\begin{subequations}\label{eomgeod}
\begin{align}
	P_{i} &= v^i \nonumber\\
	& + \frac{1}{c^2}\, \left( \frac{1}{2} v^2 v^i + 3 V v^i - 4 V_i \right) 
\nonumber\\
	& + \frac{1}{c^4}\, \left(\frac{3}{8} v^4 v^i + \frac{7}{2} V v^2 v^i \ - 
\ 4 V_j v^i v^j - 2 V_i v^2 \right. \nonumber\\
	& \qquad + \left. \frac{9}{2} V^2 v^i - 4 V V_i + 4 \hat{W}_{ij} v^j - 
8 \hat{R}_i \right) \nonumber\\
	& + \frac{1}{c^6} \left( \frac{5}{16} v^6 v^i + \frac{33}{8} V v^4 v^i - 
\frac{3}{2} V_i v^4 - 6 V_j v^i v^j v^2 + \frac{49}{4} V^2 \, v^2 v^i 
\right. \nonumber\\
 	& \qquad + 2 \hat{W}_{ij} v^j v^2 + 2 \hat{W}_{jk} v^i v^j v^k - 
10 V V_i v^2 - 20 V V_j v^i v^j \nonumber\\
 	& \qquad - 4 \hat{R}_i v^2 - 8 \hat{R}_j v^i v^j + \frac{9}{2} V^3 v^i + 
12 V_j V_j v^i + 12 \hat{W}_{ij} V v^j \nonumber\\ 
	& \qquad + 12\hat{X} v^i + 16 \hat{Z}_{ij} v^j - 10 V^2 V_i \nonumber\\
	& \qquad \left. - 8 \hat{W}_{ij} V_j - \ 8 V \hat{R}_i - 
16 \hat{Y}_i \right) \, + \calO(8)\;, \\
	F_{i} &= \partial_i V \nonumber\\ 
	& + \frac{1}{c^2} \left( - V \, \partial_i V + 
\frac{3}{2} \partial_i V \, v^2 - 4 \partial_i V_j \, v^j \right)\nonumber\\ 
	& + \frac{1}{c^4}\, \left( \frac{7}{8} \partial_i V \, v^4 - 
2 \partial_i V_j \, v^j v^2 + \frac{9}{2} V \, \partial_i V \, v^2 + 
2 \partial_i \hat{W}_{jk} \, v^j v^k - 4 V_j \, \partial_i V \, v^j 
\right. \nonumber\\ 
	& \qquad - 4 V \, \partial_i V_j \, v^j - 
\left. 8 \partial_i \hat{R}_j \, v^j + \frac{1}{2} V^2\, \partial_i V \, + 
8 V_j \, \partial_i V_j + 4 \partial_i \hat{X} \right) \nonumber\\ 
	& + \frac{1}{c^6} \left( \frac{11}{16} v^6 \partial_i V - 
\frac{3}{2} \partial_i V_j \, v^j v^4 + \frac{49}{8} V \, \partial_i V \, v^4 + 
\partial_i \hat{W}_{jk} \, v^2 v^j v^k \right. \nonumber\\ 
	& \left. \qquad - 10 V_j \, \partial_i V\, v^2 v^j - 
10 V \partial_i V_j \, v^2 v^j - 4 \partial_i \hat{R}_{j}\, v^2 v^j + 
\frac{27}{4} V^2 \,\partial_i V \, v^2 \right. \nonumber\\ 
	& \left. \qquad + 12 V_j \partial_i V_j\, v^2 + 
6 \hat{W}_{jk} \, \partial_i V\, v^j v^k + 
6 V \, \partial_i \hat{W}_{jk} \, v^j v^k + 
6 \partial_i \hat{X} v^2 \right. \nonumber\\ 
	& \left. \qquad + 8 \partial_i \hat{Z}_{jk}\, v^j v^k - 
20 V_j V \partial_i V v^j - 10 V^2 \, \partial_i V_j \, v^j - 
8 V_k \, \partial_i \hat{W}_{jk} \, v^j \right. \nonumber\\ 
	& \left. \qquad - 8 \hat{W}_{jk} \, \partial_i V_k \, v^j - 
8 \hat{R}_{j} \, \partial_i V \, v^j - 8 V \, \partial_i \hat{R}_{j}\, v^j - 
16 \partial_i \hat{Y}_j\, v^j \right. \nonumber\\ 
	& \left. \qquad - \frac{1}{6} V^3\, \partial_i V - 
4 V_j \, V_j \, \partial_i V \, + 16 \hat{R}_{j} \, \partial_i
V_j \, + 16 V_j \, \partial_i \hat{R}_j \right.\nonumber\\ 
	& \left. \qquad - 8 V\, V_j\, \partial_i V_j - 4 \hat{X}\, \partial_i V - 
4 V \, \partial_i \hat{X} \, + 
16 \partial_i \hat{T} \vphantom{\frac{11}{16}} \right) \, + \calO(8)\;.
\end{align}
\end{subequations}}\noindent
Specializing now to our problem, when computing for instance the equations of
motion for the body 1, the velocity $v^{i}$ is to be replaced by $v_{1}^{i}$
and the right-hand sides are to be evaluated at $\bfy_{1}$. However, beware
that the metric potentials are generically singular at the location of the two
particles 1 and 2 where they are meant to be evaluated. This evaluation is
thus given a sense through the Hadamard ``partie finie'' regularization
procedure explained in Ref.~\cite{BFreg}, which is of course to be performed
after computing the derivatives of the potentials. We adopt the so-called pure
Hadamard-Schwartz prescription \cite{BDE04} for the practical implementation
of this regularization. In particular, we use the distributive rule for
computing regularizations of products of potentials in Eqs.~\eqref{eomgeod},
writing
\begin{equation}\label{distributivity}
	(AB)_{1}=(A)_{1}(B)_{1}\;,
\end{equation}
for $A,B$ among (possibly derivatives of) metric potentials. The other
ingredient of this regularization is the Schwartz distributional derivative
which is used for source terms to be integrated in
Eq.~\eqref{defdalembertian}, in the form of the Gel'Fand-Shilov formula valid
for homogeneous functions, as given by
Eqs.~\eqref{gelfandshilov1}--\eqref{gelfandshilov2} below.

\begin{table}[b]
     \begin{ruledtabular}
       \caption{First line: the different spin parts ($S$) of potentials to be
         computed. Second line: the highest PN order required for the spin
         part of the potentials. Third line: for which potentials an
         \textit{all-space} (A.S.) computation of the non-compact support part
         is possible with known techniques; and, if not, which are the
         derivatives of potentials that have to be computed directly using the
         regularization in 1. In the latter case we employ the technique of
         regularized Poisson integrals described in Sec.~\ref{subsecNCS}. Note
         that all-space computations of $\hat{X}^{S}$ and $\hat{R}_{i}^{S}$ at
         the previous PN order $\calO(1)$ are possible and have been checked
         to be consistent with the regularized versions.}
	\label{tab:potentials}
     \begin{tabular}{|c|c|c|c|c|c|c|c|c|} 
		Potential & $V^{S}$ & $V_{i}^{S}$ & $\hat{X}^{S}$ & $\hat{R}_{i}^{S}$
        & $\hat{W}_{ij}^{S}$ & $\hat{T}^{S}$ & $\hat{Y}_{i}^{S}$ & 
$\hat{Z}_{ij}^{S}$ \\
	\hline
		Order & $\calO(7)$ & $\calO(5)$ & $\calO(3)$ & $\calO(3)$ & 
$\calO(3)$ & $\calO(1)$ & $\calO(1)$ & $\calO(1)$ \\
	\hline
		Computation & A.S. & A.S. & $( \partial_{i}\hat{X} )_{1}$ & 
$( \hat{R}_{i})_{1} , ( \partial_{j}\hat{R}_{i})_{1}$ & 
A.S. & $( \partial_{i}\hat{T})_{1}$ & 
$( \hat{Y}_{i})_{1} , ( \partial_{j}\hat{Y}_{i})_{1}$ & 
A.S. \\
     \end{tabular}
     \end{ruledtabular}
\end{table}
In this work, we restrict ourselves to contributions to the coordinate
acceleration that are of linear order in spins. Considering
Eq.~\eqref{eomstructure}, these linear-in-spin contributions to the geodesic
part of the acceleration can have two origins at this level: first, from the
spin contributions to the elementary metric potentials \eqref{defpotentials},
which we will have to compute at the required order, and secondly, from the
time derivative $\ud P_{i}/\ud t$, since it is understood that order-by-order
replacements of the accelerations are to be performed, which include in turn
spin contributions starting at $\calO(3)$. Table~\ref{tab:potentials}
indicates which spin parts of metric potentials are needed at which order, for
the present computation of the next-to-next-to-leading order spin-orbit
contributions to the equations of motion. As explained in
Sec.~\ref{secHadamard}, some potentials were computed in all space, but for
the most non-linear ones, only their regularized values at the location of one
of the particles could be computed. In the latter case, the calculation is
different for each different derivative structure of the potential, and
Table~\ref{tab:potentials} gives the list of these needed derivatives.

On the other hand, the effective Papapetrou part of the acceleration or force
$\calF_{i}$, which corresponds to non-geodesic motion, can be found in Paper~I
up to 2.5PN order in terms of the spin variable $S^{i}_\text{FBB}$. We give
here its complete expression up to 3.5PN order, in terms of the spin tensor
$S^{ij}$ and the various metric potentials. Defining
\begin{equation}\label{papadef}
	m \calF^{i} = \frac{1}{c^{3}}f_{3}^{i} + \frac{1}{c^{5}}f_{5}^{i} +
    \frac{1}{c^{7}}f_{7}^{i} +\calO(9) \;,
\end{equation}
and expanding the Riemann tensor in Eq.~\eqref{eomrewrite2} in terms of metric
components, themselves expanded in terms of the elementary potentials
\eqref{defpotentials}, we get:
\begin{subequations}\label{papapetroupart}
\begin{align}
f_{3}^{i} &= S^{ij}\left( \partial_{t}\partial_{j}V +
v^{k}\partial_{jk}V \right) +
S^{jk}\left( 2 v^{k}\partial_{ij}V - 2 \partial_{ij}V_{k} \right)\;,\\
f_{5}^{i} &= S^{ij} \left(- v^{k} \partial_{j}V \partial_{k}V + 
2 V v^{k} \partial_{k}\partial_{j}V + 2 \partial_{j}V_{k} \partial_{k}V + 
2 v^{j} \partial_{k}V \partial_{k}V -2 \partial_{k}V_{j} \partial_{k}V \right. 
\nonumber \\
& \left. + \partial_{j}V \partial_{t}V + 2 V \partial_{t}\partial_{j}V + 
v^{j} v^{k} \partial_{t}\partial_{k}V + v^{j} \partial_{t}^{2}V \right)
\nonumber \\
&+ S^{jk}  \left(4 v^{j} \partial_{i}V \partial_{k}V -
4 \partial_{j}V \partial_{k}V_{i} -4 \partial_{i}V \partial_{k}V_{j} + 
4 \partial_{ik}\hat{R}_{j} -4 V v^{j} \partial_{ik}V \right. \nonumber \\
& + 4 V_{j} \partial_{ik}V + 2 v^{j} v^{l} \partial_{ik}V_{l} -
2 v^{l} \partial_{ik}\hat{W}_{lj} -2 v^{j} v^{l} \partial_{kl}V_{i} + 
2 v^{l} \partial_{kl}\hat{W}_{ij} \nonumber \\ 
& \left. - 2 v^{j} \partial_{t}\partial_{i}V_{k} -
2 v^{j} \partial_{t}\partial_{k}V_{i} + 
2 \partial_{t}\partial_{k}\hat{W}_{ij}\right)\;,\\
\label{papapetroupartc}
f_{7}^{i} &= S^{ij} \left(4\partial_{j}\hat{R}_{k}\partial_{k}V -
2 V v^{k}\partial_{j}V\partial_{k}V + 4 V_{k}\partial_{j}V\partial_{k}V + 
8v^{k}\partial_{j}V_{l}\partial_{k}V_{l} + 2 V^2 v^{k}\partial_{kj}V \right. 
\nonumber \\
& + 8 v^{k} V_{l}\partial_{kj}V_{l}+ 4 v^{k}\partial_{kj}\hat{X} -
4\partial_{k}V\partial_{k}\hat{R}_{j} + 4 V v^{j}\partial_{k}V\partial_{k}V -
4V_{j}\partial_{k}V\partial_{k}V \nonumber \\
& -2 v^{k}\partial_{j}\hat{W}_{kl}\partial_{l}V + 
2 v^{j} v^{k}\partial_{k}V_{l}\partial_{l}V -
2v^{k}\partial_{k}\hat{W}_{jl}\partial_{l}V -
2 v^{j} v^{k}\partial_{l}V_{k}\partial_{l}V + 
2 v^{k}\partial_{l}\hat{W}_{kj}\partial_{l}V \nonumber \\
& + 2 V\partial_{j}V\partial_{t}V -
2 v^{k}\partial_{j}V_{k}\partial_{t}V + v^{j} v^{k}\partial_{k}V\partial_{t}V -
2 v^{k}\partial_{k}V_{j}\partial_{t}V + v^{j}\left(\partial_{t}V\right)^2 
\nonumber \\
& + 8\partial_{j}V_{k}\partial_{t}V_{k} +
4v^{j}\partial_{k}V \partial_{t}V_{k} -
2\partial_{k}V\partial_{t}\hat{W}_{jk} + 
2 V^2\partial_{t}\partial_{j}V + 8 V_{k}\partial_{t}\partial_{j}V_{k}
\nonumber \\
& \left. + 4\partial_{t}\partial_{j}\hat{X} + 
6 Vv^{j} v^{k}\partial_{t}\partial_{k}V -4 v^{k}
V_{j}\partial_{t}\partial_{k}V + 
6 V v^{j}\partial_{t}^{2}V -4 V_{j}\partial_{t}^{2}V\right) \nonumber \\
&+ S^{jk} \left(-8 V\partial_{i}V_{k}\partial_{j}V -
8\partial_{i}V\partial_{k}\hat{R}_{j} + 8 V v^{j}\partial_{i}V\partial_{k}V -
8 V_{j}\partial_{i}V\partial_{k}V -
4v^{j} v^{l}\partial_{i}V_{l}\partial_{k}V \right. \nonumber \\
& + 8\partial_{j}\hat{R}_{i}\partial_{k}V + 8 V\partial_{i}V\partial_{k}V_{j} + 
8 v^{l}\partial_{i}V_{l}\partial_{k}V_{j} + 
4v^{l}\partial_{i}V_{j}\partial_{k}V_{l} -
12 v^{j}\partial_{i}V_{l}\partial_{k}V_{l} \nonumber \\
& -4 v^{l}\partial_{j}V_{i}\partial_{k}V_{l} + 
4\partial_{j}V_{l}\partial_{k}\hat{W}_{il} + 
4\partial_{i}V_{l}\partial_{k}\hat{W}_{jl} + 
4 v^{j} v^{l}\partial_{ki}\hat{R}_{l} + 
8\hat{R}_{j}\partial_{ki}V \nonumber \\
& -4 V^2 v^{j}\partial_{ki}V + 8V V_{j}\partial_{ki}V + 
4 v^{j} v^{l} V_{l}\partial_{ki}V -8 v^{l}\hat{W}_{lj}\partial_{ki}V + 
4 V^2\partial_{ki}V_{j} \nonumber \\
& + 8v^{l} V_{l}\partial_{ki}V_{j} + 8 V v^{j} v^{l}\partial_{ki}V_{l} -
16 v^{j} V_{l}\partial_{ki}V_{l} + 4\hat{W}_{jl}\partial_{ki}V_{l} -
4 V v^{l}\partial_{ki}\hat{W}_{lj} \nonumber \\
& + 4 V_{l}\partial_{ki}\hat{W}_{lj} -8 v^{j}\partial_{ki}\hat{X} + 
8\partial_{ki}\hat{Y}_{j}-8 v^{l}\partial_{ki}\hat{Z}_{lj} -
4 v^{j} v^{l}\partial_{i}V_{k}\partial_{l}V \nonumber \\
& + 4 v^{j} v^{l}\partial_{k}V\partial_{l}V_{i} -
8 v^{l}\partial_{k}V_{j}\partial_{l}V_{i} + 
4v^{j} v^{l}\partial_{i}V\partial_{l}V_{k} -
4 v^{l}\partial_{i}V_{j}\partial_{l}V_{k} -
4 v^{j}\partial_{i}V_{l}\partial_{l}V_{k} \nonumber \\
& + 4 v^{l}\partial_{j}V_{i}\partial_{l}V_{k} -
4\partial_{j}V_{l}\partial_{l}\hat{W}_{ik} -
4 v^{j} v^{l}\partial_{lk}\hat{R}_{i} + 
4 v^{l}\hat{W}_{ij}\partial_{lk}V -8 V v^{j} v^{l}\partial_{lk}V_{i} \nonumber
\\
& -8 v^{l} V_{i}\partial_{lk}V_{j} + 4 V v^{l}\partial_{lk}\hat{W}_{ij} + 
8 v^{l}\partial_{lk}\hat{Z}_{ij} + 4
v^{j}\partial_{i}\hat{W}_{kl}\partial_{l}V + 
4 v^{j}\partial_{k}\hat{W}_{il}\partial_{l}V \nonumber \\
& -4 v^{j}\partial_{l}\hat{W}_{ik}\partial_{l}V -
4 v^{j}\partial_{k}V_{l}\partial_{l}V_{i} + 
4 v^{j}\partial_{l}V_{k}\partial_{l}V_{i} -
4\partial_{i}\hat{W}_{kl}\partial_{l}V_{j} -
4\partial_{k}\hat{W}_{il}\partial_{l}V_{j} \nonumber \\
& + 4\partial_{l}\hat{W}_{ik}\partial_{l}V_{j} + 
4 v^{j}\partial_{k}V\partial_{t}V_{i} -8\partial_{k}V_{j}\partial_{t}V_{i} + 
4 v^{j}\partial_{i}V\partial_{t}V_{k} + 
8\partial_{j}V_{i}\partial_{t}V_{k} \nonumber \\
& -4\partial_{j}V\partial_{t}\hat{W}_{ik} -
4 v^{j}\partial_{t}\partial_{i}\hat{R}_{k} + 
4 v^{j} V_{k}\partial_{t}\partial_{i}V -
8 V v^{j}\partial_{t}\partial_{i}V_{k} +
8 V_{j}\partial_{t}\partial_{i}V_{k} \nonumber \\
& + 2 v^{j} v^{l}\partial_{t}\partial_{i}\hat{W}_{lk} -
4 v^{j}\partial_{t}\partial_{k}\hat{R}_{i} + 
4\hat{W}_{ij}\partial_{t}\partial_{k}V -
8 V v^{j}\partial_{t}\partial_{k}V_{i} -
8 V_{i}\partial_{t}\partial_{k}V_{j} \nonumber \\
& \left. + 4 V\partial_{t}\partial_{k}\hat{W}_{ij} + 
8\partial_{t}\partial_{k}\hat{Z}_{ij} -
2 v^{j} v^{l}\partial_{t}\partial_{l}\hat{W}_{ik} -
2 v^{j}\partial_{t}^{2}\hat{W}_{ik}\right)\;.
\end{align}
\end{subequations}
If we are computing the Papapetrou acceleration of body 1, $\calF_{1}^{i}$, we
have to replace in the latter expressions $S^{ij}$ by $S_{1}^{ij}$ and $v^{i}$
by $v_{1}^{i}$. Because of the explicit spin factor, at linear order in spins
only the non-spin parts of the potentials are needed, and $S_{2}^{ij}$ does
not appear in this part of the acceleration. The factor $m\equiv m_1$ in
Eq.~\eqref{papadef} is at the origin of all terms depending on the spins by
unit mass, \textit{i.e.} through $S_1^{ij}/m_1$ or $S_2^{ij}/m_2$ in the final
results. Note that terms with two velocities appear in
Eqs.~\eqref{papapetroupart} because of the replacement of $S_{1,2}^{0i}$ by
$S_{1,2}^{ij}$ according to \eqref{S0itoSij}. Again all products of singular
potentials have to be regularized following the rule of the pure
Hadamard-Schwartz regularization, notably the distributivity rule
\eqref{distributivity}.\footnote{Note that we checked that, at this order, it
  makes no difference in the final result to evaluate the product of
  derivatives of potentials before or after taking the partie finie at 1.}
However we shall find in Sec.~\ref{secDimreg} that the computation of the
value at 1 of one particular potential, namely $\partial_{jk}\hat{Y}_{i}$,
which is especially singular, \textit{a priori} requires the use of
dimensional regularization.

As we said, since the Papapetrou part of the acceleration already includes an
explicit spin factor, only the non-spin parts of the potentials are required
there, and most of these have already been computed in previous works such as
\cite{BFP98,BFeom}. Notable exceptions are the second spatial derivatives of
the potentials $\hat{R}_{i}$ and $\hat{Y}_{i}$, which are defined in
  Eqs.~\eqref{defpotentials}; we address the computation of these new
regularized potentials in Secs.~\ref{secHadamard} and \ref{secDimreg}.


\section{Hadamard regularization of potentials}
\label{secHadamard}

\subsection{Notation and general points}
\label{subsecGeneral}

Throughout the rest of this paper, we use the following notation. The
trajectories of the two bodies, with masses $m_1$ and $m_2$ and spin tensors
$S_1^{ij}$ and $S_2^{ij}$, are denoted $\bfy_{1}(t)$ and $\bfy_{2}(t)$, and
their coordinate velocities $\bfv_{1}(t)$ and $\bfv_{2}(t)$. The inter-body
distance is $r_{12} = |\bfy_{1}-\bfy_{2}|$, and we set $n_{12}^{i} =
(y_{1}^{i}-y_{2}^{i})/|\bfy_{1}-\bfy_{2}|$. For a generic field point $\bfx$,
we pose $r_{1}=|\bfx-\bfy_{1}|$, $n_{1}^{i} =
(x^{i}-y_{1}^{i})/|\bfx-\bfy_{1}|$ and similarly for 2. We denote by
$\partial_{i}$, $\partial^{1}_{i}$ and $\partial^{2}_{i}$ the partial
derivative with respect to $x^{i}$, $y_{1}^{i}$ and $y_{2}^{i}$. The symbol $1
\leftrightarrow 2$ means the expression obtained by the exchange of the two
particles. Lengthy calculations are performed with the help of the scientific
software Mathematica{\footnotesize \textregistered}, supplemented by the
package xTensor~\cite{xtensor} dedicated to tensor calculus.

We systematically expand the retardations inside the retarded integral
\eqref{defdalembertian}, truncated to the appropriate order. For instance,
truncating at $\calO(3)$, we have:
\begin{align}\label{taylorexpand}
	-4\pi\,\Box_{\mathcal{R}}^{-1}f(\bfx,t) &= 
\int \frac{\ud^{3}\bfx'}{|\bfx-\bfx'|}f(\bfx',t) - 
\frac{1}{c} \int \ud^{3}\bfx' \partial_{t}f(\bfx',t) + 
\frac{1}{2c^{2}} \int \ud^{3}\bfx' 
|\bfx-\bfx'| \partial_{t}^{2}f(\bfx',t) \nonumber \\
	& \quad + \calO(3)\; .
\end{align}
In this expression, time derivatives may be pulled out of the integrals, and
it is understood that accelerations and time derivatives of the spin are to be
replaced by the already computed lower order equations of motion and spin
precession equations. Note that the second term in this expansion has no
dependence to the field point $\bfx$.

Next, notice that we are using Dirac delta functions in the stress-energy
momentum tensor \eqref{Tmunu3D}, while it makes no sense as a distribution
\textit{\`a la} Schwartz when acting on the class of general functions in the
problem, which are generically singular at the locations of the two particles.
We refer to \cite{BFeom} for a complete discussion of this issue, which was
dealt with by defining a class of ``pseudo-functions'' and using Hadamard
regularization to define the value of a singular function at one of its singular
points. In our case, we will give a sense to Dirac delta functions inside
integrals, following the rule
\begin{subequations}\label{rulesdirac}
\begin{equation}\label{rulesdirac1}
	\int \ud^{3}\mathbf{x} F(\mathbf{x})\delta_{1} = (F)_{1}  \;, 
\end{equation}
where $\delta_{1}=\delta(\mathbf{x}-\mathbf{y}_{1})$ and $(F)_{1}$ means the
Hadamard ``partie finie'' of $F$ at the point $\mathbf{y}_{1}$ \cite{BFreg}.
This rule is extended in an obvious way to derivatives of delta functions, for
instance
\begin{equation}\label{rulesdirac2}
	\int \ud^{3}\mathbf{x} F(\mathbf{x})\partial_{i}\delta_{1}  = 
-(\partial_{i}F)_{1} \;.
\end{equation}
\end{subequations}

Other issues are the problem of distributivity of the partie finie already
mentionned, and the treatment of the distributional parts of derivatives. Here
we shall not use the ``extended Hadamard regularization'' introduced in
Ref.~\cite{BFreg}; instead we will apply primarily the ``pure
Hadamard-Schwartz'' prescription described in Ref.~\cite{BDE04} which is
sufficient for most of our computations. The pure Hadamard-Schwartz
regularization constitutes the core of the most powerful and fundamental
regularization procedure which is dimensional regularization
and has been successfully applied to the problem of the equations of motion in
Refs.~\cite{DJSdim,BDE04}. Thus, in principle, the result of the pure
Hadamard-Schwartz regularization of Poisson-type integrals is to be
supplemented by a contribution from dimensional regularization which appears
when the integral develops a pole $\propto 1/(d-3)$ in the dimension. The pole
corresponds to the appearance of logarithmic divergences in the Hadamard
regularization (see Sec.~\ref{subsecNCS}). Here, we stick to the pure
Hadamard-Schwartz prescription for the computation of the spin parts of
potentials, where no problematic logarithms appear and where we expect that
there is no difference between the extended Hadamard or pure Hadamard-Schwartz
regularizations and the dimensional regularization (as it was the case up to
2.5PN order in the non-spin equations of motion \cite{BFP98}). On the other
hand, as already mentioned we do use dimensional regularization for one
particular potential in the Papapetrou part of the equations of motion, as
discussed in Sec.~\ref{secDimreg}.

For the distributional part of derivatives of singular functions, we employ a
particular form of the Gel'Fand-Shilov formula valid for homogeneous
functions. Denoting by $\partial_{i}$ the full derivative, including the
distributional part, and by $\partial_{i}^\text{ord}$ the ordinary part, which
also acts on the Dirac delta functions, we get, in the relevant cases of
double derivatives and a simple function of type $n^L/r^m$ (which is
homogeneous of degree $-m$),
\begin{equation}\label{gelfandshilov1}
	\partial_{ij} f = \partial_{ij}^\text{ord}f
    +D_{i}\bigl[\partial_{j}^\text{ord}f\bigr]+\partial_{i}^\text{ord}D_{j}[f]\;,
\end{equation}
where the distributional part is given when $\ell+m$ is an odd integer by
\begin{equation}\label{gelfandshilov2}
	D_{i}\left( \frac{n^{L}}{r^{m}} \right) = 
4\pi \frac{(-)^{m}2^{m}(\ell+1)!(\frac{\ell+m-1}{2})!}{(\ell+m)!}
\sum_{p=p_{0}}^{[m/2]} \frac{\Delta^{p-1}\partial_{(M-2P}\, 
\delta_{iL+2P-M)}}{2^{2p}(p-1)!(m-2p)!(\frac{\ell+1-m}{2}+p)!}  \;,
\end{equation}
and is zero when $\ell+m$ is even.\footnote{Here we pose
  $p_{0}=\text{Max}[1,(m-\ell-1)/2]$, and $[m/2]$ means the integer part.
  Notation for multi-indices is for instance $L=j_1 j_2\cdots j_\ell$,
  \textit{i.e.} $l$ is the number of indices on $n^L=n^{j_1}\cdots n^{j_\ell}$.
  We denote $\delta_{2K}=\delta_{j_1j_2}\cdots\delta_{j_{2k-1}j_{2k}}$ where
  $2k=\ell+2p-m+1$ is the number of indices in the multi-index $iL+2P-M$. The
  parenthesis refer to the complete symmetrization of indices.}
In the present context we apply the formulae
\eqref{gelfandshilov1}--\eqref{gelfandshilov2} to the expansion of a singular
function $f$ around the two singularities $\bfy_{1}$ or $\bfy_{2}$.
 
\subsection{Leading-order potentials}
\label{subsecLO}

We first deal with the lowest-order potentials, namely the computation of the
leading spin-orbit part of the potentials $V$, $V_{i}$ and $\hat{W}_{ij}$. We
give all the results for completeness, translating results from Paper~I in
terms of the spin tensor $S^{ij}$. By expliciting the expression of the
dipolar part \eqref{Tmunudipolar} of the stress-energy tensor, one may check
that the leading order for the spin part of the sources $\sigma$,
$\sigma_{i}$, and $\sigma_{ij}$ is:
\begin{subequations}\label{sigmaLO}
\begin{align}
	\sigma^{S} & = \frac{2}{c^{3}} S_{1}^{ij} v_{1}^{i}\partial_{j}\delta_{1}
    + 
1\leftrightarrow2 +\calO(5) \; ,\\ 
	\sigma_{i}^{S} & = \frac{1}{2c} S_{1}^{ij} \partial_{j}\delta_{1} + 
1\leftrightarrow2 +\calO(3)\; , \\ 
	\sigma_{ij}^{S} & = -\frac{1}{c}
    S_{1}^{k(i}v_{1}^{j)} \partial_{k}\delta_{1} + 
1\leftrightarrow2 +\calO(3) \; , 
\end{align}
\end{subequations}
where $(ij)$ indicates symmetrization, $\delta_{1}$
stands for $\delta(\mathbf{x}-\mathbf{y}_{1})$, and the gradients are taken
with respect to the field point $\mathbf{x}$. Given these expressions for the
sources of potentials, and using $\Delta (1/r_{1})=-4\pi\delta_{1}$ (with
$r_{1}=|\mathbf{x}-\mathbf{y}_{1}|$), Eqs.~\eqref{sigmaLO} then yield:
\begin{subequations}\label{potentialsLO}
\begin{align}
	V^{S} & = \frac{2G}{c^{3}} S_{1}^{ij} v_{1}^{i}\partial_{j}
\left(\frac{1}{r_{1}}\right) + 1\leftrightarrow2 +\calO(5) \; ,\\ 
	V_{i}^{S} & = \frac{G}{2c} S_{1}^{ij} \partial_{j}
\left(\frac{1}{r_{1}}\right) + 1\leftrightarrow2 +\calO(3) \; ,\\ 
	\hat{W}_{ij}^{S} & = -\frac{G}{c} \left( S_{1}^{k(i}v_{1}^{j)} - 
\delta^{ij}S_{1}^{kl}v_{1}^{l}\right) \partial_{k}
\left(\frac{1}{r_{1}}\right)+ 1\leftrightarrow2 +\calO(3) \; .
\end{align}
\end{subequations}
Notice that the leading-order of the spin part of these potentials has the
typical dipolar structure $\partial(1/r)$, while it is simply $1/r$
(monopolar) for the non-spin part. Note also the important fact that $V^{S}$
starts at $\calO(3)$ and has no $\calO(1)$ contribution. This makes the
structure of the spin part of potentials different from the one of the
non-spin part, and shifts a number of contributions to higher PN order. In
particular, the non-compact support part of $\hat{W}_{ij}^{S}$, coming from
$\Box_{\mathcal{R}}^{-1}(-\partial_{i}V\partial_{j}V)$, starts only at
$\calO(3)$ and not $\calO(1)$. This in turn implies that the leading order of
$\hat{W}_{ij}^{S}$ has the simpler structure of a compact-supported term,
which makes a number of sources of higher potentials less non-linear, such as
the non-compact-supported term $\Box_{\mathcal{R}}^{-1}(
\hat{W}_{ij} \partial^2_{ij} V)$ in $\hat{X}^{S}$.

\subsection{Higher order compact-support terms}
\label{subsecCS}

We turn now to the higher-order PN computation of the compact-supported parts
of potentials, \textit{i.e.} whose sources include $\sigma$, $\sigma_{i}$ or
$\sigma_{ij}$. We will take the potentials $V^{S}$ as an example and give only
a schematic view of these computations, since their implementation is
relatively straightforward. By expliciting the expression \eqref{Tmunudipolar}
of the stress-energy tensor, we get the following structure for the
compact-supported sources:
\begin{equation}\label{sigmaSstruct}
	\sigma_{1}(\bfx,t) =\tilde{\mu}_{1\text{M}} \,\delta_{1} + 
\frac{1}{\sqrt{-g(\bfx)}} \tilde{\mu}_{1\text{D}} \,\delta_{1} + 
\frac{1}{\sqrt{-g(\bfx)}} \partial_{t}\bigl(\nu_{1\text{D}}\,\delta_{1}\bigr) + 
\frac{1}{\sqrt{-g(\bfx)}} \partial_{i}(\nu_{1\text{D}}^i\,\delta_{1}) \;,
\end{equation}
indifferently for bodies 1 and 2, where the subscripts M and D refer to the
monopolar and dipolar parts of the stress-energy tensor. The quantity
$\tilde{\mu}_{1\text{M}}$ already intervenes in the study of the equations of
motion without spin, and its 3.5PN expression in terms of the metric
potentials is given in Eq.~(4.3) of Ref.~\cite{NB05}. Next, we Taylor expand
the retardations \eqref{taylorexpand} for each term. The time derivatives are
pulled out of the integrals, which are then evaluated using the rules
\eqref{rulesdirac} for the delta functions and their derivatives, keeping in
mind that the factors $1/\sqrt{-g}$ in \eqref{sigmaSstruct} depend on the
field point $\bfx$.

As a check of the method, we also computed all the compact-support parts of
the spin potentials using a fully non-distributive prescription, \textit{i.e.}
keeping the dependence in the field point $\bfx$ as long as possible (also
when evaluating the quantities such as $\nu_{1\text{D}}$), and taking the
Hadamard partie finie at the very end of the calculation. We obtained no
difference with the pure Hadamard distributive prescription
\eqref{distributivity}.

Notice that $V^{S}$, for instance, gets contributions from both the dipolar
and monopolar parts of the stress-energy tensor. For the dipolar part, the
quantities $\tilde{\mu}_{1\text{D}}$, $\nu_{1\text{D}}$ and
$\nu_{1\text{D}}^i$ all display an explicit spin factor. For the monopolar
part, spin terms appear indirectly in two ways: through the spin
contributions to the metric potentials in the expression of
$\tilde{\mu}_{1\text{M}}$, and through the acceleration replacements when
evaluating time derivatives.

\subsection{Non-compact support terms}
\label{subsecNCS}

The non-compact supported potentials are more complicated, since
their source is itself made of potentials and does not allow the simple
integration procedure with delta functions. However, if the source has a simple
structure, some integrations can be computed explicitly to obtain a solution
valid for a generic field point $\bfx$. One encounters often the source
structure $\partial(1/r)\partial^2(1/r)$, which arises at leading-order in all
$\partial A^{NS}\partial B^{S}$ terms, where $A$ and $B$ are
chosen among $\{V, V_{i}\}$ and $\{V,V_{i,}\hat{W}_{ij}\}$ respectively, as
can be checked from the leading-order structure \eqref{potentialsLO}. Using
the same techniques as in previous works (\textit{e.g.}~\cite{BFP98,BFeom}), in
particular using the function $g$ such that
\begin{subequations}\label{defg}
\begin{align}
	\Delta g & = \frac{1}{r_{1}r_{2}} \;, \\ 
	g & \equiv \ln (r_{1}+r_{2}+r_{12})\; , 
\end{align}
\end{subequations}
we find the following relations:
\begin{subequations}\label{inverselaplacians}
\begin{align}
	\Delta^{-1} \left[ \partial_{i}\left( \frac{1}{r_{1}}\right)
\partial_{jk}\left( \frac{1}{r_{1}}\right) \right] & = 
\frac{1}{16}\left[ \partial_{ijk}\ln(r_{1})+
\left( \delta^{ij}\partial_{k}+\delta^{ik}\partial_{j}-
\delta^{jk}\partial_{i} \right)\left(\frac{1}{r_{1}^{2}}\right) \right] \;, \\ 
	\Delta^{-1} \left[ \partial_{i}\left( \frac{1}{r_{1}}\right)
\partial_{jk}\left( \frac{1}{r_{2}}\right) \right] & = -
\partial^{1}_{i}\partial^{2}_{jk} g\;,
\end{align}
\end{subequations}
where for instance $\partial^{1}_{i}=\partial/\partial y_{1}^{i}$. With these
relations in hand, completing the results for the compact-support parts, we
are able to straightforwardly compute the leading order $\calO(1)$ of
$\hat{X}^{S}$, $\hat{R}_{i}^{S}$, $\hat{Z}_{ij}^{S}$, as well as the 1PN
relative order $\calO(3)$ of $\hat{W}_{ij}^{S}$, as indicated in
Table~\ref{tab:potentials}.

We now turn to the computation of the partie finie of Poisson integrals, of
the type of the first and third term in \eqref{taylorexpand}.\footnote{Pure
  spatial integrals like the second term of \eqref{taylorexpand}, where the
  integrand is independent of $\bfx$, are computed using the Hadamard
  prescription for singular integrals, following the procedure described in
  Ref.~\cite{BFreg}.} When direct integration is not possible (at least using
known results) for the non-compact support part of non-linear potentials, one
by-passes the difficulty by directly computing the Hadamard regularized value
of the integral at the points $\bfy_{1}$ or $\bfy_{2}$. By doing so, one will
not have access to the full information about the potentials (hence the
metric) everywhere in space, but only the information relevant for the
computation of the equations of motion. Since the computation is different for
each derivative of the potential, one has to figure out which derivatives of
which potential will be needed in the final computation.
Table~\ref{tab:potentials} already gave the required spin parts of potentials.
Furthermore we find that two new non-spin (\textit{NS}) potentials, featuring
two derivatives, have to be computed in addition to known previous results
\cite{BFeom}:
\begin{equation}\label{NSpotentials}
	\bigl(\partial_{jk}\hat{R}_{i}^{NS}\bigr)_{1} ~\text{to order
      $\calO(2)$} 
\;, ~\text{and}~ \bigl(\partial_{jk}\hat{Y}_{i}^{NS}\bigr)_{1} 
~\text{to order $\calO(0)$} \;.
\end{equation}

Following Ref.~\cite{BFreg} (to which we refer for details and definitions),
we define generic Poisson integrals and twice-iterated Poisson integrals:
\begin{subequations}\label{PQdef}
\begin{align}
	P(\bfx) & = -\frac{1}{4\pi} \text{Pf} \int 
\frac{\ud^{3}\bfx'}{|\bfx-\bfx'|}F(\bfx') \;, \label{Pdef} \\
	Q(\bfx) & = -\frac{1}{4\pi} \text{Pf} \int 
\ud^{3}\bfx' |\bfx-\bfx'|F(\bfx') \;, \label{Qdef}
\end{align}
\end{subequations}
where the source $F$ is singular at $\bfy_{1}$ and $\bfy_{2}$. The symbol $\text{Pf}$
stands for the Hadamard partie finie for integrals diverging near the
particle locations. This regularization procedure consists in removing from the
integration domain two spherical balls of radius $s$ surrounding the two
singularities, substracting the purely divergent part, and taking the limit
$s\to 0$. The partie finie regularization depends on two arbitrary constants
$s_1$ and $s_2$ (see Ref.  \cite{BFreg}).
The results for the extended
Hadamard partie finie (in the sense of Eqs.~(5.3)-(5.4) in \cite{BFreg}) of
the Poisson potentials $P$ and $Q$ are:
\begin{subequations}\label{PQres}
\begin{align}
	(P)_{1} & = -\frac{1}{4\pi} \text{Pf} \int 
\frac{\ud^{3}\bfx}{r_{1}}F(\bfx)+
\left[\ln\left( \frac{r'_{1}}{s_{1}}\right)-1\right] (r_{1}^{2}F)_{1} \;, \\
	(Q)_{1} & = -\frac{1}{4\pi} \text{Pf} \int 
\ud^{3}\bfx \, r_{1}F(\bfx)+
\left[\ln\left( \frac{r'_{1}}{s_{1}}\right) + \frac{1}{2}\right] 
(r_{1}^{4}F)_{1} \;, \\
	(\partial_{i}P)_{1} & = -\frac{1}{4\pi} \text{Pf} \int 
\ud^{3}\bfx \frac{{n_{1}}^{i}}{{r_{1}}^{2}}F(\bfx)+
\ln\left( \frac{r'_{1}}{s_{1}}\right) (r_{1} n_{1}^{i} F)_{1} \;, \\
	(\partial_{i}Q)_{1} & = \frac{1}{4\pi} \text{Pf} \int 
\ud^{3}\bfx \, {n_{1}}^{i}F(\bfx)-\left[\ln\left( \frac{r'_{1}}{s_{1}}\right) -
\frac{1}{2}\right]  (r_{1}^{3} n_{1}^{i} F)_{1} \;.
\end{align}
\end{subequations}
In these formulae, the first term would correspond to the naive prescription
of taking directly $\bfx=\bfy_{1}$ inside the integrals \eqref{PQdef}, and the
second term features a regularization constant $\ln(r'_{1})$ coming from the
singular limit $\bfx'\to\bfy_{1}$. Notice that the constant $s_{1}$
automatically cancels between the two terms, while the $\text{Pf}$ term may
induce a dependence on $s_{2}$, so that the result depends on two constants
$r'_{1}$ and $s_2$. For the present work we had to extend these formulae to
the case of a double gradient, in order to compute the potentials
\eqref{NSpotentials}. Using the same method as in Ref.~\cite{BFreg}, we
obtained the corresponding required expressions for the double derivative:
\begin{subequations}\label{dijPQres}
\begin{align}
	(\partial_{ij}P)_{1} & = -\frac{1}{4\pi} \text{Pf} \int 
\ud^{3}\bfx \frac{3{n_{1}}^{ij}-\delta^{ij}}{{r_{1}}^{3}}F(\bfx) + 
\ln\left( \frac{r'_{1}}{s_{1}}\right) 
\left( (3 n_{1}^{ij}-\delta^{ij}) F\right)_{1} +
\frac{\delta^{ij}}{3} (F)_{1} \;, \label{dijPres} \\
	(\partial_{ij}Q)_{1} & =  -\frac{1}{4\pi} \text{Pf} \int 
\ud^{3}\bfx \frac{\delta^{ij}-{n_{1}}^{ij}}{r_{1}}F(\bfx) + 
\left[\ln\left( \frac{r'_{1}}{s_{1}}\right)+\frac{1}{2}\right]
 \left( (\delta^{ij}-n_{1}^{ij})r_{1}^{2} F\right)_{1} -
\delta^{ij} (r_{1}^{2}F)_{1} \;. \label{dijQres}
\end{align}
\end{subequations}
One readily checks that these expressions yield the correct results when
contracted with $\delta^{ij}$: namely $(\Delta P)_{1} = (F)_{1}$ and $(\Delta
Q)_{1} = 2 (P)_{1}$. Crucial for this check, notice the last term in
Eq.~\eqref{dijPres} which stems from a distributional contribution obtained
when evaluating the two derivatives under the integral \eqref{Pdef}. It can
be absorbed by replacing, in the first term of
\eqref{dijPres}, the ordinary factor
$(3{n'}_{1}^{ij}-\delta^{ij})/{r'}_{1}^{3}$ by the distributional derivative
$\partial'_{ij}(1/r'_{1})$.

With these tools in hand, we were able to compute all the needed non-compact
support parts of potentials. Importantly, we get no contribution of the
singular parts proportional to $\ln(r'_{1})$ in Eqs.~\eqref{PQres}, in any of
the spin parts of potentials listed in Table~\ref{tab:potentials}. This gives
a strong indication that the Hadamard regularization (actually the pure
Hadamard-Schwartz version of it) is sufficient to deal with all spin parts of
potentials. This is not surprising because the calculation we are doing is
only of 2PN relative order, while past experience in the non-spin case
\cite{BFP98,BFeom} says that Hadamard's regularization fails no earlier than
at 3PN relative order.


\section{Dimensional regularization for one potential}
\label{secDimreg}

By contrast, we have found that in the Papapetrou part of the acceleration,
which is specific to the spin-case (and outside past experience), we do get a
contribution proportional to $\ln(r'_{1})$ in one of the new evaluations we
had to perform, namely when computing the double-derivative potential
$(\partial_{j}\partial_{k}\hat{Y}_{i}^{NS})_{1}$ at Newtonian order using
Eq.~\eqref{dijPres}. All the other \textit{NS} potentials, including the other
new one $(\partial_{j}\partial_{k}\hat{R}_{i}^{NS})_{1}$ we had to compute,
could be safely obtained with the Hadamard regularization of
Sec.~\ref{secHadamard}.

The appearance of this $\ln(r'_{1})$ tells us, on the contrary, that
Hadamard's regularization is insufficient for the potential
$(\partial_{j}\partial_{k}\hat{Y}_{i}^{NS})_{1}$, and that we \textit{a
  priori} need dimensional regularization in order to get rid of possible
ambiguities in the final equations of motion, as was the case of the non-spin
3PN equations of motion in \cite{BFeom,BDE04}. On the other hand, when
considering directly the Papapetrou contribution to the acceleration, we see
that the dangerous contribution exists in only one term, with index structure
$S_{1}^{jk}(\partial_{i}\partial_{j}\hat{Y}_{k}^{NS})_{1}$, and we find that
the problematic logarithms $\ln(r'_{1})$ cancel out in the final result
because of the antisymmetry of $S_{1}^{jk}$. Therefore, we expect beforehand
the pole $\propto (d-3)^{-1}$ we shall obtain in dimensional
regularization to actually vanish in the final acceleration. Nevertheless,
even though the result will be pole-free, we know that the finite part
$\propto (d-3)^{0}$ should play a crucial role, and therefore we must
\textit{a priori} apply the powerful but tedious procedure of dimensional
regularization to the problematic potential
$(\partial_{j}\partial_{k}\hat{Y}_{i}^{NS})_{1}$.

We refer to \cite{BDE04} for precise definitions and technical details about
the method, which consists in studying the problem in $d$ spatial dimensions,
treating $d$ as a complex variable, and taking the analytical continuation of
the obtained results when $d\rightarrow 3$. We will associate a superscript
$(d)$ to quantites which are defined in this $d$ dimensional setting. One
considers a class of functions regular everywhere except at the points
$\bfy_{1}$ and $\bfy_{2}$, and admitting expansions of the type ($\forall
N\in\mathbb{N}$):
\begin{equation}\label{Fddev}
	F^{(d)}(\mathbf{x})=
\sum_{\substack{p_0\leqslant p\leqslant N\\ q_0\leqslant q\leqslant q_1}}
    r_1^{p+q\varepsilon}\mathop{f}_1{}_{p,q}^{(\varepsilon)}(\mathbf{n}_1)+o(r_1^N)
    \;, 
\end{equation}
and similarly around $\bfy_{2}$, with $\varepsilon \equiv d-3$, and $q_{0}$,
$q_{1}$, $p_{0}\in \mathbb{Z}$. As long as there are no terms in these
expansions with both $p<0$ and $q=0$, and that there is no angular dependence
of the $p=0,q=0$ term in the expansion, which is always the case in practice,
the presence of the factor $r_{1}^{q\varepsilon}$ allows one to write, by
application of the analytical continuation on $\varepsilon$:
\begin{subequations}
\begin{align}
	(F^{(d)}G^{(d)})(\bfy_{1}) &= F^{(d)}(\bfy_{1})G^{(d)}(\bfy_{1}) \;, \\
	F^{(d)}(\bfx)\delta^{(d)}(\bfx-\bfy_{1}) &= 
F^{(d)}(\bfy_{1})\delta^{(d)}(\bfx-\bfy_{1}) \;,
\end{align}
\end{subequations}
which justifies the use of distributivity in our previous Hadamard
three-dimensional computations. Indeed, it was checked in previous work
\cite{BDE04} on the
3PN non-spin equations of motion that the pure Hadamard
prescription agreed with the $d\rightarrow 3$ limit of the dimensional
prescription for all the terms except those developing a pole. As explained in
Sec.~\ref{subsecCS}, although we did not perform a full dimensional
regularization computation of all quantities to be evaluated at $\bfy_{1}$ or
$\bfy_{2}$, we checked that distributive and fully non-distributive
prescriptions for the computation of compact-supported potentials yielded the
same final result, which gives us confidence in this distributive
prescription.\footnote{Within the dimensional regularization scheme,
  distributional derivatives can be treated as Schwartz distributional
  derivatives, \textit{e.g.} using Gel'Fand-Shilov formulae in $d$
  dimensions.}

The treatment of Poisson integrals such as \eqref{PQdef} goes as follows.
Defining $P^{(d)}(\bfx)$ as:
\begin{equation}\label{Pdimreg}
	P^{(d)}(\bfx) = -\frac{\tilde{k}}{4\pi} \int 
\frac{\ud^{d}\bfx'}{|\bfx-\bfx'|^{d-2}}F^{(d)}(\bfx') \;, 
\end{equation}
where $\tilde{k} \equiv \Gamma\left( \frac{d-2}{2}
\right)/\pi^{\frac{d-2}{2}}$, analytical continuation allows us to take
derivatives under the integral, to set apart the distributional contribution,
and to make directly the replacement $\bfx \rightarrow \bfy_{1}$ under the
remaining integral. We obtain:
\begin{equation}\label{dijPdimreg}
	(\partial_{ij}P^{(d)})(\bfy_{1}) = -\frac{\tilde{k}}{4\pi}(d-2) \int 
\ud^{d}\bfx \,\frac{d \, n_{1}^{ij} - \delta^{ij}}{r_{1}^{d}} F^{(d)}(\bfx) + 
\frac{\delta^{ij}}{d}F^{(d)}(\bfy_{1}) \;,
\end{equation}
where the second term on the right-hand side corresponds to the distributional
contribution. It reads $F^{(d)}(\bfy_{1}) \equiv {}_1f_{0,0}^{(\varepsilon)}$,
according to the expansion \eqref{Fddev},
and must not have any angular dependence (since dimensional regularization
does not include angular averaging in its definition, unlike Hadamard's partie
finie). We set here all the terms $r_{1}^{q\varepsilon}
{}_1f_{0,q}^{(\varepsilon)}(\mathbf{n}_{1})$ with $q\neq 0$ to $0$ since they are cancelled by
analytical continuation (and have in general an angular dependence). We also
assume that ${}_1f_{0,0}^{(\varepsilon)}$ has no pole in $\varepsilon$ which
is true, since no poles are generated at the level of the sources of
potentials for this calculation.

When $d \rightarrow 3$ we have $F^{(d)} \rightarrow F$, where $F$ is the
corresponding function in the Hadamard $3$-dimensional context. That function
admits the expansion
\begin{equation}\label{Fdev}
	F(\mathbf{x})=\sum_{p_0\leqslant p\leqslant N} 
r_{1}^{p}\mathop{f}_1{}_{p}(\mathbf{n}_1)+ o(r_1^N) \;.
\end{equation}
The Hadamard partie finie of the function is defined by the angular average
over the unit direction $\mathbf{n}_1$ of the term $p=0$, say
\begin{equation}\label{F1dev}
	 (F)_{1}=\bigl< \mathop{f}_1{}_{0}(\mathbf{n}_1) \bigr> \;.
\end{equation}
The $3$-dimensional coefficients ${}_1f_{p}(\mathbf{n}_1)$ are related to the
$\varepsilon \rightarrow 0$ limits of the $d$-dimensional coefficients
${}_1f_{p,q}^{(\varepsilon)}(\mathbf{n}_1)$ appearing in Eq.~\eqref{Fddev} by
\begin{equation}\label{relf}
	 \mathop{f}_{1}{}_{p}(\mathbf{n}_1) = 
\sum_{q_{0} \leqslant q \leqslant q_{1}}  \mathop{f}_{1}{}^{(0)}_{p,q} (\mathbf{n}_1)\;,
\end{equation}
which in turn gives us the link between the analytical continuation when
$d\rightarrow 3$ of $F^{(d)}(\bfy_{1})$ and the Hadamard partie finie
$(F)_{1}$:
\begin{equation}\label{}
	 (F)_{1} = F^{(3)}(\bfy_{1}) + 
\sum_{\substack{ q_{0} \leqslant q \leqslant q_{1} \\ q \neq 0}} 
\bigl< \mathop{f}_{1}{}^{(0)}_{0,q} (\mathbf{n}_1) \bigr> \;.
\end{equation}

Next, we address the link between the Hadamard-regularized Poisson integral
\eqref{PQdef} and its dimensional regularization version \eqref{Pdimreg}.
Defining:
\begin{equation}\label{DP1}
	\mathcal{D}(\partial_{ij}P)(1) \equiv 
(\partial_{ij}P^{(d)})(\mathbf{y}_1)-(\partial_{ij}P)_1\;,
\end{equation}
and following the same steps as in Ref.~\cite{BDE04}, 
we obtain, working at the zeroth order in $\varepsilon$,
\begin{align}\label{DdijP1}
	\mathcal{D}(\partial_{ij}P)(1) &= -\frac{1}{\varepsilon} 
\sum_{q_0\leqslant q\leqslant q_1}\left(\frac{1}{q}+ 
\varepsilon\ln r_1'\right)\bigl< (d\,n_{1}^{ij} - \delta^{ij})
 \mathop{f}_1{}_{0,q}^{(\varepsilon)} \bigr> \nonumber \\
	& -\frac{1}{\varepsilon (1+\varepsilon)}
\sum_{q_0\leqslant q\leqslant q_1}\left(\frac{1}{q+1}+
\varepsilon\ln s_2\right) 
\sum_{\ell=0}^{+\infty}\frac{(-)^\ell}{\ell!}\partial^{1}_{ijL} 
\left(\frac{1}{r_{12}^{1+\varepsilon}}\right)
\bigl< n_2^L\mathop{f}_2{}_{-\ell-3,q}^{(\varepsilon)}\bigr> \nonumber \\
	& - \frac{\delta_{ij}}{3} \sum_{\substack{ q_{0} \leqslant q 
\leqslant q_{1} \\ 
q \neq 0}} 
\bigl< \mathop{f}_{1}{}^{(0)}_{0,q} (\mathbf{n}_1) \bigr> + \mathcal{O}(\varepsilon)\,.
\end{align}

The result of the PN iteration of the metric, starting with the
$d+1$-dimensional Einstein equations, and the corresponding definitions for
the metric potentials are given in Sec.~II of Ref.~\cite{BDE04}.
In particular, the $d$-dimensional definition of the dangerous potential
$\hat{Y}_{i}$ is given by (2.12f) there.\footnote{Notice that because we have
  already seen that the pole $1/\varepsilon$ will necessarily cancel out in
  the final result, we do not have to worry about extending the Papapetrou
  part of the force \eqref{papapetroupart} to $d$ dimensions. \textit{I.e.},
  we can evaluate the term $\partial_{jk}\hat{Y}_{i}$ assuming it carries the
  3-dimensional coefficient, which is $+8$ in this case, see one of the
    terms in Eq.~\eqref{papapetroupartc}.} We applied the previous method to
the computation of the quantity $\mathcal{D}(\partial_{jk}\hat{Y}_{i})(1)$
defined above. This required computing the singular expansion of the needed
non-compact support sources in $d$ dimensions up to the order $p=0$, which was
done following the methods of Ref.~\cite{BDE04}. Our result for the pole part
is quite compact,
\begin{equation}\label{pole}
	\mathcal{D}(\partial_{jk}\hat{Y}_{i})(1) = 
\frac{1}{\varepsilon} \frac{G^{3}m_{1}^{2}m_{2}}{252} v_{12}^{l} 
\partial^{1}_{ijkl}\left( \frac{1}{r_{12}} \right) + \mathcal{O}(\varepsilon^0)\;.
\end{equation}
Since this expression is symmetric by exchange of the two indices $i$ and $j$
(or $k$), it doesn't contribute to the final equations of motion where only
the contraction $S_{1}^{jk}(\partial_{ij}\hat{Y}_{k}^{NS})_{1}$ is
involved. This confirms that the final equations of motion in dimensional
regularization are directly pole-free. Furthermore, we also checked that the
finite part $\propto\varepsilon^0$ coming from Eq.~\eqref{DdijP1} also cancels
out in the final result (\textit{i.e.} after contraction with the spin tensor
$S_{1}^{jk}$). Although the latter fact could not \textit{a priori} be guessed
beforehand, it implies that finally the dimensional regularization was
superfluous for the computation of the dangerous term
$S_{1}^{jk}(\partial_{ij}\hat{Y}_{k}^{NS})_{1}$. This result gives us
further confidence that the pure Hadamard-Schwartz regularization is
sufficient for all our calculations.


\section{Results for the equations of motion and precession}
\label{secResults}

Since higher-order results take the form of quite long formulae with many
similar terms, we tried to adopt a systematic presentation. First, we split
the expressions in terms of powers of $G$ and according to the powers of the
masses $m_{1}$ and $m_{2}$. Each spin-dependent term is then denoted using the
convention that $\alpha^{i}_{p,q}$ gathers all the terms featuring
$G^{p+q}m_{1}^{p}m_{2}^{q}$. For any two vectors $\mathbf{a}$ and
$\mathbf{b}$, we use the notation $(ab)$ for the scalar product,
\textit{i.e.} $(ab)\equiv \mathbf{a}\cdot\mathbf{b}=a^ib^i$, and we define
$(Sab)\equiv S^{ij}a^{i}b^{j}$.

Before presenting the results, let us address an important point: every time
an expression has at least one free index, it might admit several equivalent
writings, in terms of the vectors appearing in the problem and of the spin
tensors. Indeed, any of the final results, which are functions of time only
(such as the acceleration of one of the bodies or the time derivative of one
of the spins), can be written with the three vectors $n_{12}^{i}$,
$v_{1}^{i}$, $v_{2}^{i}$, and with at most one occurence of either one of the
spin tensors $S_{1}^{ij}$ and $S_{2}^{ij}$ since we are working at linear
order in spins. As the spatial indices run on three different values, we have
the two identities:
\begin{subequations}
\begin{align}
	S^{[ij}a^{k}b^{l]}a_{j}b_{k}c_{l} &= 0 \;, \label{dimidentity1} \\
	a_{m}S^{m[i}a^{j}b^{k}c^{l]}a_{j}b_{k}c_{l} &= 0 \;, \label{dimidentity2}
\end{align}
\end{subequations}
where the brackets indicate antisymmetrization over the indices, where
$S^{ij}$ is one of the spin tensors and where $(a,b,c)$ is a permutation of
the set of three vectors $(n_{12}, v_{1}, v_{2})$. The number of identities of
this kind that one must take into account depends on the number of vectors and
tensors at disposal, and on the number of free indices.
As an example, \eqref{dimidentity1} with $S=S_{1}$, $a=n_{12}$, $b=v_{1}$ and
$c=v_{2}$ gives, once expanded:
\begin{align}
	0 = & - S_{1}^{ij}n_{12}^{j} v_{1}^{2} (n_{12}v_{2}) + 
S_{1}^{ij}n_{12}^{j} (n_{12}v_{1}) (v_{1}v_{2}) + 
n_{12}^{i} (S_{1}n_{12}v_{1}) (v_{1}v_{2})  \nonumber\\
	& - n_{12}^{i} v_{1}^{2} (S_{1}n_{12}v_{2}) + 
n_{12}^{i} (n_{12}v_{1}) (S_{1}v_{1}v_{2}) - 
v_{1}^{i} (S_{1}n_{12}v_{1}) (n_{12}v_{2})  \nonumber\\
	& + v_{1}^{i} (n_{12}v_{1}) (S_{1}n_{12}v_{2}) - 
v_{1}^{i} (S_{1}v_{1}v_{2}) + 
S_{1}^{ij} v_{1}^{j} (n_{12}v_{1}) (n_{12}v_{2})  \nonumber\\
	& - S_{1}^{ij} v_{1}^{j}(v_{1}v_{2})  - 
S_{1}^{ij} v_{2}^{j}(n_{12}v_{1})^2 + S_{1}^{ij} v_{2}^{j} v_{1}^{2}  \;.
\end{align}
Hence, one must keep in mind that there is no unique writing for the results
we are going to present, and take this into account when comparing two
expressions. This becomes particularly important when using the method of
undetermined coefficients (for instance when looking for a contact
transformation between harmonic and ADM variables, see Sec.~\ref{subsecADM}):
the system of independent equations that these coefficients have to solve is
to be determined only after taking into account the complete list of these
identities. Notice also that the use of an antisymmetric spin tensor reduces
the number of these identities compared to the use of a vector spin variable
and a Levi-Civita symbol, which was one of our motivations for changing the
spin variable with respect to Paper~I. A straightforward but cumbersome
work around of this problem when comparing results is to project everything in
an arbitrary orthonormal basis.

\subsection{Spin evolution equation}
\label{subsecSpinevol}

First, we give the spin evolution equation, \textit{i.e.} the equation giving
the time derivative of the spatial components of the spin tensor $S^{ij}$, up
to 2PN order. This result was in fact already contained in Paper~I, under the
form of a precession equation for the spin vector $\mathbf{S}_{\text{FBB}}$,
and it is a mere matter of traduction between the spin variables. Beware that
our spin tensor is not of conserved norm, $S^{ij}S^{ij}\neq \text{const}$
beyond leading order. This equation is needed each time we perform
order-by-order reduction when evaluating time derivatives.

Since we are working at 2PN relative order, it is sufficient to know the 2PN
or $\calO(4)$ spin evolution equation. This also means that the amount of
non-linearity in this computation is less that in the computation of the
acceleration. Indeed, expliciting Eq.~\eqref{papprecessionlinear} in terms of
potentials, we see that we need only 2PN potentials,
and we can ignore the spin contributions in these potentials, working at
linear order in spins. However, notice that, since the leading-order spin
contributions to the total angular momentum of the system is of the form
$\mathbf{S}_{1}/c+\mathbf{S}_{2}/c$, with $\mathbf{S}_{1,2}$ spin vectors, the
order of this spin evolution equation required for finding a conserved total
angular momentum at 3.5PN order is not 2PN but 3PN.

Defining
\begin{equation}\label{dtS1struct}
\frac{\ud S_1^{ij}}{\ud t} = \frac{1}{c^2}B_\mathrm{1PN}^{ij} + \frac{1}{c^4}
B_\mathrm{2PN}^{ij} + \mathcal{O}\left(6\right) \, ,
\end{equation}
we obtain for the 1PN order:
\begin{equation}\label{dtS12}
B_\mathrm{1PN}^{ij} = \frac{G m_{2}}{r_{12}^{2}}\left[
  2S_{1}^{ij}(n_{12}v_{12}) + 
4 n_{12}^{[i}S_{1}^{j]k}v_{12}^{k} -2 v_{1}^{[i}S_{1}^{j]k}n_{12}^{k} + 
4 v_{2}^{[i}S_{1}^{j]k}n_{12}^{k} \right] \, ,
\end{equation}
and for the 2PN order, splitting the result in terms of powers of $G$ and
occurence of the masses as explained above:
\begin{equation}\label{dtS14struct}
B_\mathrm{2PN}^{ij} = \frac{G}{r_{12}^{2}} \beta^{ij}_{0,1} m_{2} +
\frac{G^{2}}{r_{12}^{3}} \left[ \beta^{ij}_{1,1} m_{1}m_{2} + \beta^{ij}_{0,2}
  m_{2}^{2} \right] \;,
\end{equation}
where
\begin{subequations}
\begin{align}
    \beta^{ij}_{0,1} &= n_{12}^{[i}S_{1}^{j]k}v_{12}^{k} \left[ -6 (n_{12}v_{2})^2 - 
4 (v_{12}v_{2}) \right] + 3 (n_{12}v_{2})^2 v_{12}^{[i}S_{1}^{j]k}n_{12}^{k} + 
2 (n_{12}v_{2}) v_{12}^{[i}S_{1}^{j]k}v_{12}^{k} \nonumber \\
& + v_{2}^{[i}S_{1}^{j]k}n_{12}^{k} \left[ -3 (n_{12}v_{2})^2 - 
4 (v_{12}v_{2}) \right] + v_{2}^{[i}S_{1}^{j]k}v_{12}^{k} \left[ 4 (n_{12}v_{12}) + 
2 (n_{12}v_{2}) \right] \nonumber \\
& + S_{1}^{ij} \left[ -3 (n_{12}v_{12}) (n_{12}v_{2})^2 + 2 (n_{12}v_{2})
  (v_{12}v_{2}) \right] \;,\\
    \beta^{ij}_{1,1} &= 32 (n_{12}v_{12}) n_{12}^{[i}S_{1}^{j]k}n_{12}^{k} -
    14 n_{12}^{[i}S_{1}^{j]k}v_{12}^{k} - 12 v_{12}^{[i}S_{1}^{j]k}n_{12}^{k}
    \nonumber \\
& - 2 v_{2}^{[i}S_{1}^{j]k}n_{12}^{k} + S_{1}^{ij} \left[ 2 (n_{12}v_{12}) + 2
  (n_{12}v_{2}) \right] \;,\\
    \beta^{ij}_{0,2} &= -4 (n_{12}v_{12}) n_{12}^{[i}S_{1}^{j]k}n_{12}^{k} + 2
    v_{12}^{[i}S_{1}^{j]k}n_{12}^{k} - 2 (n_{12}v_{12}) S_{1}^{ij} \;.
\end{align}
\end{subequations}
Note that, with the series of potentials already computed for the
    3PN equations of motion without spins \cite{BF00,BDE04}, we are able to
  control the precession equations up to the next 3PN order; this will be
  investigated in future work.

\subsection{Acceleration}
\label{subsecAcc}

The spin contributions in the acceleration have the following structure, with
our PN counting valid for maximally spinning-objects:
\begin{align}\label{a1struct}
\frac{\ud v_1^i}{\ud t} &=
A^i_\mathrm{N}+\frac{1}{c^2}A^i_\mathrm{1PN}+\frac{1}{c^3}
\mathop{A^i}_{S}{}_{\!\mathrm{1.5PN}}
+\frac{1}{c^4}\left[A^i_\mathrm{2PN}+
\mathop{A^i}_{SS}{}_{\!\mathrm{2PN}}\right]
+\frac{1}{c^5}\left[A^i_\mathrm{2.5PN}+
\mathop{A^i}_{S}{}_{\!\mathrm{2.5PN}}\right] \nonumber \\
& \qquad + \frac{1}{c^6}\left[A^i_\mathrm{3PN}+
\mathop{A^i}_{SS}{}_{\!\mathrm{3PN}}\right] +
\frac{1}{c^7}\left[A^i_\mathrm{3.5PN}+
  \mathop{A^i}_{S}{}_{\!\mathrm{3.5PN}}\right]  +\calO(8) \, ,
\end{align}
where the \textit{S} subscript indicates the spin-orbit contributions, and the
\textit{SS} subscript indicates contributions that are quadratic in the spins
and which we neglect in this work.\footnote{Notice that, when doing the
  calculation in the original harmonic coordinates, one actually obtains some
  3PN spin-orbit contributions. However, these are pure gauge, as was already
  explained in Appendix~A of Ref.~\cite{BBF11}, and are eliminated by the
  gauge transformation $x^{\mu}\rightarrow x^{\mu}+\delta X^{\mu}$ with
  $\delta X^{0}=0$ and
\begin{equation*}
	\delta X^{i} = - \frac{G^{2}}{r_{12}^{2}c^{6}}\left(m_{1}S_{2}^{ij} -
      m_{2}S_{1}^{ij} \right) n_{12}^{j} \,,
\end{equation*}
whose effect on the accelerations is $\delta a_{1}^{i} = \delta a_{2}^{i} =
\ud^{2}\delta X^{i}/\ud t^{2} + \calO(8)$. This gauge transformation obviously
respects the harmonicity condition in a perturbative sense, since $\Box \delta
X^{\mu} = \calO(8)$.}
For the leading-order spin contributions, we get
\label{a1istruct}
\begin{align}
	 m_{1}\mathop{A^i}_{S}{}_{\!\mathrm{1.5PN}} & = 
\frac{G}{r_{12}^{3}}\left[m_{2}\left( 3 S_{1}^{ij}n_{12}^{j} (n_{12}v_{12}) + 
6 (S_{1}n_{12}v_{12})n_{12}^{i} -3 S_{1}^{ij}v_{12}^{j} \right) \right. \nonumber \\
	 & \qquad \quad \left. + m_{1} \left( 6 S_{2}^{ij}n_{12}^{j} (n_{12}v_{12}) + 
6 (S_{2}n_{12}v_{12})n_{12}^{i} - 4 S_{2}^{ij}v_{12}^{j} \right) \right] \;.
\end{align}
It depends on the two velocities $v^i_{1}$ and $v^i_{2}$ only through the
relative velocity $v_{12}^i=v_{1}^i-v_{2}^i$. This is imposed by the Lorentz
invariance of the harmonic-coordinate equations of motion, which reduces to a
Galilean invariance at leading order (see Sec.~\ref{subsecLorentz}). For
the 1PN relative order contributions, we get the following:
\begin{equation}\label{a1iS5}
	m_{1} \mathop{A^i}_{S}{}_{\!\mathrm{2.5PN}} =  
\frac{G}{r_{12}^{3}}\left[ \alpha^{i}_{0,1} m_{2} + 
\alpha^{i}_{1,0} m_{1} \right] + 
\frac{G^{2}}{r_{12}^{4}} \left[ \alpha^{i}_{0,2} m_{2}^{2}  + 
\alpha^{i}_{1,1} m_{1}m_{2} + \alpha^{i}_{2,0} m_{1}^{2} \right]\, ,
\end{equation}
where the coefficients are completely equivalent to those obtained in Paper~I
and read\footnote{The link between the spin variable used in Paper~I and the
  present spin tensor is provided in Appendix~\ref{appSiSij}.}
\allowdisplaybreaks{
\begin{subequations}
\begin{align}
    \alpha^{i}_{0,1} &= S_{1}^{ij}n_{12}^{j} 
\left[ -\frac{15}{2} (n_{12}v_{12}) (n_{12}v_{2})^2 - 
\frac{3}{2} (n_{12}v_{12}) v_{12}^{2} - 
3 (n_{12}v_{12}) (v_{12}v_{2})  \right. \nonumber \\
& \qquad \qquad \quad \left. + 3 (n_{12}v_{2}) (v_{12}v_{2}) - 
\frac{3}{2} (n_{12}v_{12}) v_{2}^{2} \right] \nonumber \\
& + S_{1}^{ij}v_{12}^{j} \left[ -3 (n_{12}v_{12}) (n_{12}v_{2}) + 
\frac{9}{2} (n_{12}v_{2})^2 + \frac{3}{2} v_{12}^{2} + 6 (v_{12}v_{2}) + 
\frac{3}{2} v_{2}^{2} \right] \nonumber \\
& + (S_{1}n_{12}v_{12})n_{12}^{i} \left[ -15 (n_{12}v_{2})^2 - 
3 v_{12}^{2} - 12 (v_{12}v_{2}) - 3 v_{2}^{2} \right] \nonumber \\
& + (S_{1}n_{12}v_{12})v_{12}^{i} \left[ -3 (n_{12}v_{12}) - 
6 (n_{12}v_{2}) \right] + 3 (n_{12}v_{12}) (S_{1}n_{12}v_{12})v_{2}^{i} \nonumber \\
& + 3 (n_{12}v_{12}) (S_{1}n_{12}v_{2})v_{12}^{i} + 
3 (n_{12}v_{12}) (S_{1}n_{12}v_{2})v_{2}^{i} \nonumber \\
& - 3 (S_{1}v_{12}v_{2})v_{12}^{i} - 3 (S_{1}v_{12}v_{2})v_{2}^{i} \;,\\
    \alpha^{i}_{1,0} &= S_{2}^{ij}n_{12}^{j} 
\left[ -15 (n_{12}v_{12}) (n_{12}v_{2})^2 - 6 (n_{12}v_{12}) (v_{12}v_{2}) + 
6 (n_{12}v_{2}) (v_{12}v_{2}) - 3 (n_{12}v_{12}) v_{2}^{2} \right] \nonumber \\
& + S_{2}^{ij}v_{12}^{j} \left[ 6 (n_{12}v_{2})^2 + 4 (v_{12}v_{2}) + 
2 v_{2}^{2} \right] + (S_{2}n_{12}v_{12})n_{12}^{i} 
\left[ -15 (n_{12}v_{2})^2 - 
6 (v_{12}v_{2}) - 3 v_{2}^{2} \right] \nonumber \\
& + (S_{2}n_{12}v_{12})v_{12}^{i} \left[ -6 (n_{12}v_{12}) - 
6 (n_{12}v_{2}) \right] + 6 (n_{12}v_{12}) (S_{2}n_{12}v_{2})v_{12}^{i} \nonumber \\
& + 6 (n_{12}v_{12}) (S_{2}n_{12}v_{2})v_{2}^{i} - 
4 (S_{2}v_{12}v_{2})v_{12}^{i} - 4 (S_{2}v_{12}v_{2})v_{2}^{i} \;,\\
	\alpha^{i}_{0,2} &= -6 (n_{12}v_{12}) S_{1}^{ij}n_{12}^{j} + 
6 S_{1}^{ij}v_{12}^{j} - 12 (S_{1}n_{12}v_{12})n_{12}^{i} \;,\\
    \alpha^{i}_{1,1} &= -14 (n_{12}v_{12}) S_{1}^{ij}n_{12}^{j} + 
14 S_{1}^{ij}v_{12}^{j} - 26 (S_{1}n_{12}v_{12})n_{12}^{i} \nonumber \\
& -16 (n_{12}v_{12}) S_{2}^{ij}n_{12}^{j} + 12 S_{2}^{ij}v_{12}^{j} - 
20 (S_{2}n_{12}v_{12})n_{12}^{i} \;,\\
	\alpha^{i}_{2,0} &= S_{2}^{ij}n_{12}^{j} \left[ -\frac{31}{2} (n_{12}v_{12}) + 
2 (n_{12}v_{2}) \right] + \frac{23}{2} S_{2}^{ij}v_{12}^{j} - 
\frac{45}{2} (S_{2}n_{12}v_{12})n_{12}^{i}\;.
\end{align}
\end{subequations}}
Finally, the main result of our work, namely the next-to-next-to-leading 3.5PN
spin-orbit contributions to the acceleration, reads:
\begin{align}\label{a1iS7}
	m_{1}\mathop{A}_{S}{}^i_{\!\mathrm{3.5PN}} =&  
\frac{G}{r_{12}^{3}}\left[ \gamma^{i}_{0,1} m_{2} + 
\gamma^{i}_{1,0} m_{1} \right] + 
\frac{G^{2}}{r_{12}^{4}} \left[ \gamma^{i}_{0,2} m_{2}^{2} + 
\gamma^{i}_{1,1} m_{1}m_{2} + \gamma^{i}_{2,0} m_{1}^{2} \right] \nonumber \\
	& + \frac{G^{3}}{r_{12}^{5}} \left[ \gamma^{i}_{0,3} m_{2}^{3} + 
\gamma^{i}_{1,2}  m_{1}m_{2}^{2} + \gamma^{i}_{2,1}  m_{1}^{2}m_{2} + 
\gamma^{i}_{3,0}  m_{1}^{3} \right]\;,
\end{align}
where
\allowdisplaybreaks{
\begin{subequations}
\begin{align}
    \gamma^{i}_{0,1} &= S_{1}^{ij}n_{12}^{j}
 \left[ \frac{105}{8} (n_{12}v_{12}) (n_{12}v_{2})^4 + 
\frac{15}{4} (n_{12}v_{12}) (n_{12}v_{2})^2 v_{12}^{2} - 
\frac{3}{8} (n_{12}v_{12}) v_{12}^{4}  \right. \nonumber \\
& \qquad \qquad \; \; \left.  + 
\frac{15}{2} (n_{12}v_{12}) (n_{12}v_{2})^2 (v_{12}v_{2}) - 
\frac{15}{2} (n_{12}v_{2})^3 (v_{12}v_{2}) - 
\frac{3}{2} (n_{12}v_{12}) v_{12}^{2} (v_{12}v_{2})  \right. \nonumber \\
& \qquad \qquad \; \; \left.  - 
\frac{3}{2} (n_{12}v_{2}) v_{12}^{2} (v_{12}v_{2}) - 
\frac{3}{2} (n_{12}v_{12}) (v_{12}v_{2})^2 - 
3 (n_{12}v_{2}) (v_{12}v_{2})^2  \right. \nonumber \\
& \qquad \qquad \; \; \left.  - 
\frac{15}{4} (n_{12}v_{12}) (n_{12}v_{2})^2 v_{2}^{2} - 
\frac{3}{4} (n_{12}v_{12}) v_{12}^{2} v_{2}^{2} - 
\frac{3}{2} (n_{12}v_{12}) (v_{12}v_{2}) v_{2}^{2}  \right. \nonumber \\
& \qquad \qquad \; \; \left.  + 
\frac{3}{2} (n_{12}v_{2}) (v_{12}v_{2}) v_{2}^{2} - 
\frac{3}{8} (n_{12}v_{12}) v_{2}^{4} \right] \nonumber \\
& + S_{1}^{ij}v_{12}^{j} 
\left[ \frac{15}{2} (n_{12}v_{12}) (n_{12}v_{2})^3 - 
\frac{45}{8} (n_{12}v_{2})^4 + 
\frac{3}{2} (n_{12}v_{12}) (n_{12}v_{2}) v_{12}^{2} - 
\frac{9}{4} (n_{12}v_{2})^2 v_{12}^{2}  \right. \nonumber \\
& \qquad \qquad \; \; \left.  + 
\frac{3}{8} v_{12}^{4} + 
3 (n_{12}v_{12}) (n_{12}v_{2}) (v_{12}v_{2}) - 
12 (n_{12}v_{2})^2 (v_{12}v_{2}) - 
\frac{3}{2} (v_{12}v_{2})^2  \right. \nonumber \\
& \qquad \qquad \; \; \left.  - 
\frac{3}{2} (n_{12}v_{12}) (n_{12}v_{2}) v_{2}^{2} + 
\frac{9}{4} (n_{12}v_{2})^2 v_{2}^{2} + 
\frac{3}{4} v_{12}^{2} v_{2}^{2} + 3 (v_{12}v_{2}) v_{2}^{2} + 
\frac{3}{8} v_{2}^{4} \right] \nonumber \\
& + (S_{1}n_{12}v_{12})n_{12}^{i} 
\left[ \frac{105}{4} (n_{12}v_{2})^4 + 
\frac{15}{2} (n_{12}v_{2})^2 v_{12}^{2} - 
\frac{3}{4} v_{12}^{4} + 30 (n_{12}v_{2})^2 (v_{12}v_{2}) + 
3 (v_{12}v_{2})^2  \right. \nonumber \\
& \qquad \qquad \qquad \quad \; \; \left.  - 
\frac{15}{2} (n_{12}v_{2})^2 v_{2}^{2} - 
\frac{3}{2} v_{12}^{2} v_{2}^{2} - 6 (v_{12}v_{2}) v_{2}^{2} - 
\frac{3}{4} v_{2}^{4} \right] \nonumber \\
& + (S_{1}n_{12}v_{12})v_{12}^{i} 
\left[ \frac{15}{2} (n_{12}v_{12}) (n_{12}v_{2})^2 + 
15 (n_{12}v_{2})^3 + \frac{3}{2} (n_{12}v_{12}) v_{12}^{2} + 
3 (n_{12}v_{2}) v_{12}^{2} \right. \nonumber \\
& \qquad \qquad \qquad \quad \; \; \left.  + 
3 (n_{12}v_{12}) (v_{12}v_{2}) + 9 (n_{12}v_{2}) (v_{12}v_{2}) - 
\frac{9}{2} (n_{12}v_{12}) v_{2}^{2} - 
3 (n_{12}v_{2}) v_{2}^{2} \right] \nonumber \\
& + (S_{1}n_{12}v_{12})v_{2}^{i} 
\left[ -\frac{15}{2} (n_{12}v_{12}) (n_{12}v_{2})^2 - 
\frac{3}{2} (n_{12}v_{12}) v_{12}^{2} - 
3 (n_{12}v_{12}) (v_{12}v_{2}) \right. \nonumber \\
& \qquad \qquad \qquad \quad \; \left.  + 
3 (n_{12}v_{2}) (v_{12}v_{2}) - 
\frac{3}{2} (n_{12}v_{12}) v_{2}^{2} \right] \nonumber \\
& + (S_{1}n_{12}v_{2})v_{12}^{i} 
\left[ -\frac{15}{2} (n_{12}v_{12}) (n_{12}v_{2})^2 - 
\frac{3}{2} (n_{12}v_{12}) v_{12}^{2} - 
3 (n_{12}v_{12}) (v_{12}v_{2})\right. \nonumber \\
& \qquad \qquad \qquad \quad \; \left.  + 
3 (n_{12}v_{2}) (v_{12}v_{2}) - 
\frac{3}{2} (n_{12}v_{12}) v_{2}^{2} \right] \nonumber \\
& + (S_{1}n_{12}v_{2})v_{2}^{i} 
\left[ -\frac{15}{2} (n_{12}v_{12}) (n_{12}v_{2})^2 - 
\frac{3}{2} (n_{12}v_{12}) v_{12}^{2} -
 3 (n_{12}v_{12}) (v_{12}v_{2})\right. \nonumber \\
& \qquad \qquad \qquad \quad \left.  + 
3 (n_{12}v_{2}) (v_{12}v_{2}) - 
\frac{3}{2} (n_{12}v_{12}) v_{2}^{2} \right] \nonumber \\
& + (S_{1}v_{12}v_{2})v_{12}^{i} \left[ -3 (n_{12}v_{12}) (n_{12}v_{2}) + 
\frac{9}{2} (n_{12}v_{2})^2 + \frac{3}{2} v_{12}^{2} + 
6 (v_{12}v_{2}) + \frac{3}{2} v_{2}^{2} \right] \nonumber \\
& + (S_{1}v_{12}v_{2})v_{2}^{i}
 \left[ -3 (n_{12}v_{12}) (n_{12}v_{2}) + \frac{9}{2} (n_{12}v_{2})^2 + 
\frac{3}{2} v_{12}^{2} + 6 (v_{12}v_{2}) + \frac{3}{2} v_{2}^{2} \right] \;,\\
   \gamma^{i}_{1,0} &= S_{2}^{ij}n_{12}^{j} 
\left[ \frac{105}{4} (n_{12}v_{12}) (n_{12}v_{2})^4 + 
15 (n_{12}v_{12}) (n_{12}v_{2})^2 (v_{12}v_{2}) -
15 (n_{12}v_{2})^3 (v_{12}v_{2})  \right. \nonumber \\
& \qquad \qquad \quad \left.  - 6 (n_{12}v_{2}) (v_{12}v_{2})^2 - 
\frac{15}{2} (n_{12}v_{12}) (n_{12}v_{2})^2 v_{2}^{2} - 
3 (n_{12}v_{12}) (v_{12}v_{2}) v_{2}^{2}  \right. \nonumber \\
& \qquad \qquad \quad \left.  + 
3 (n_{12}v_{2}) (v_{12}v_{2}) v_{2}^{2} - 
\frac{3}{4} (n_{12}v_{12}) v_{2}^{4} \right] \nonumber \\
& + S_{2}^{ij}v_{12}^{j} \left[ -\frac{15}{2} (n_{12}v_{2})^4 - 
6 (n_{12}v_{2})^2 (v_{12}v_{2}) + 3 (n_{12}v_{2})^2 v_{2}^{2} + 
2 (v_{12}v_{2}) v_{2}^{2} + \frac{1}{2} v_{2}^{4} \right] \nonumber \\
& + (S_{2}n_{12}v_{12})n_{12}^{i}
\left[ \frac{105}{4} (n_{12}v_{2})^4 + 
15 (n_{12}v_{2})^2 (v_{12}v_{2}) - 
\frac{15}{2} (n_{12}v_{2})^2 v_{2}^{2} - 
3 (v_{12}v_{2}) v_{2}^{2} - \frac{3}{4} v_{2}^{4} \right] \nonumber \\
& + (S_{2}n_{12}v_{12})v_{12}^{i} 
\left[ 15 (n_{12}v_{12}) (n_{12}v_{2})^2 + 15 (n_{12}v_{2})^3 - 
3 (n_{12}v_{12}) v_{2}^{2} - 3 (n_{12}v_{2}) v_{2}^{2} \right] \nonumber \\
& + (S_{2}n_{12}v_{2})v_{12}^{i}
\left[ -15 (n_{12}v_{12}) (n_{12}v_{2})^2 - 6 (n_{12}v_{12}) (v_{12}v_{2}) + 6
  (n_{12}v_{2}) (v_{12}v_{2}) - 
3 (n_{12}v_{12}) v_{2}^{2} \right] \nonumber \\
& + (S_{2}n_{12}v_{2})v_{2}^{i} 
\left[ -15 (n_{12}v_{12}) (n_{12}v_{2})^2 - 
6 (n_{12}v_{12}) (v_{12}v_{2}) + 6 (n_{12}v_{2}) (v_{12}v_{2}) - 
3 (n_{12}v_{12}) v_{2}^{2} \right] \nonumber \\
& + (S_{2}v_{12}v_{2})v_{12}^{i} \left[ 6 (n_{12}v_{2})^2 + 
4 (v_{12}v_{2}) + 2 v_{2}^{2} \right] \nonumber \\
& + (S_{2}v_{12}v_{2})v_{2}^{i} \left[ 6 (n_{12}v_{2})^2 + 
4 (v_{12}v_{2}) + 2 v_{2}^{2} \right] \;,\\
    \gamma^{i}_{0,2} &= S_{1}^{ij}n_{12}^{j} 
\left[ 18 (n_{12}v_{12}) (n_{12}v_{2})^2 - 
3 (n_{12}v_{12}) v_{12}^{2} + 
6 (n_{12}v_{12}) (v_{12}v_{2})  \right. \nonumber \\
& \qquad \qquad \; \; \left.  - 
6 (n_{12}v_{2}) (v_{12}v_{2}) + 
3 (n_{12}v_{12}) v_{2}^{2} \right] \nonumber \\
& + S_{1}^{ij}v_{12}^{j} \left[ -
6 (n_{12}v_{12}) (n_{12}v_{2}) - 12 (n_{12}v_{2})^2 + 
3 v_{12}^{2} - 3 v_{2}^{2} \right] \nonumber \\
& + (S_{1}n_{12}v_{12})n_{12}^{i} \left[ 36 (n_{12}v_{2})^2 - 
6 v_{12}^{2} + 6 v_{2}^{2} \right] + 
(S_{1}n_{12}v_{12})v_{12}^{i} \left[ -6 (n_{12}v_{12}) + 
12 (n_{12}v_{2}) \right] \nonumber \\
& + 6 (n_{12}v_{12}) (S_{1}n_{12}v_{12})v_{2}^{i} - 
6 (n_{12}v_{12}) (S_{1}n_{12}v_{2})v_{12}^{i} - 
6 (n_{12}v_{12}) (S_{1}n_{12}v_{2})v_{2}^{i} \nonumber \\
& + 6 (S_{1}v_{12}v_{2})v_{12}^{i} + 6 (S_{1}v_{12}v_{2})v_{2}^{i} \;,\\
   \gamma^{i}_{1,1} &= S_{1}^{ij}n_{12}^{j} 
\left[ -\frac{375}{2} (n_{12}v_{12})^3 - 
33 (n_{12}v_{12})^2 (n_{12}v_{2}) + 
45 (n_{12}v_{12}) (n_{12}v_{2})^2 + 
\frac{177}{2} (n_{12}v_{12}) v_{12}^{2}  \right. \nonumber \\
& \qquad \qquad \quad \left.  + 14 (n_{12}v_{12}) (v_{12}v_{2}) - 
14 (n_{12}v_{2}) (v_{12}v_{2}) + 
7 (n_{12}v_{12}) v_{2}^{2} \right] \nonumber \\
& + S_{1}^{ij}v_{12}^{j} \left[ \frac{271}{2} (n_{12}v_{12})^2 + 
39 (n_{12}v_{12}) (n_{12}v_{2}) - 32 (n_{12}v_{2})^2 - 
\frac{73}{2} v_{12}^{2} - 28 (v_{12}v_{2}) - 7 v_{2}^{2} \right] \nonumber \\
& + (S_{1}n_{12}v_{12})n_{12}^{i} 
\left[ -\frac{663}{2} (n_{12}v_{12})^2 - 
42 (n_{12}v_{12}) (n_{12}v_{2}) + 90 (n_{12}v_{2})^2 + 
\frac{135}{2} v_{12}^{2}  \right. \nonumber \\
& \qquad \qquad \qquad \quad \; \;  \left.  + 52 (v_{12}v_{2}) + 
13 v_{2}^{2} \right] \nonumber \\
& + (S_{1}n_{12}v_{12})v_{12}^{i} \left[ 122 (n_{12}v_{12}) + 
20 (n_{12}v_{2}) \right] - 
14 (n_{12}v_{12}) (S_{1}n_{12}v_{12})v_{2}^{i} \nonumber \\
& - 14 (n_{12}v_{12}) (S_{1}n_{12}v_{2})v_{12}^{i} - 
14 (n_{12}v_{12}) (S_{1}n_{12}v_{2})v_{2}^{i} + 
14 (S_{1}v_{12}v_{2})v_{12}^{i} + 14 (S_{1}v_{12}v_{2})v_{2}^{i} \nonumber \\
& +  S_{2}^{ij}n_{12}^{j} \left[ 48 (n_{12}v_{12}) (n_{12}v_{2})^2 + 
16 (n_{12}v_{12}) (v_{12}v_{2}) - 16 (n_{12}v_{2}) (v_{12}v_{2}) + 
8 (n_{12}v_{12}) v_{2}^{2} \right] \nonumber \\
& + S_{2}^{ij}v_{12}^{j} \left[ -24 (n_{12}v_{2})^2 - 12 (v_{12}v_{2}) - 
6 v_{2}^{2} \right] \nonumber \\
& + (S_{2}n_{12}v_{12})n_{12}^{i} \left[ 60 (n_{12}v_{2})^2 + 
20 (v_{12}v_{2}) + 10 v_{2}^{2} \right] \nonumber \\
& + 20 (n_{12}v_{2}) (S_{2}n_{12}v_{12})v_{12}^{i} - 
16 (n_{12}v_{12}) (S_{2}n_{12}v_{2})v_{12}^{i} - 
16 (n_{12}v_{12}) (S_{2}n_{12}v_{2})v_{2}^{i} \nonumber \\
& + 12 (S_{2}v_{12}v_{2})v_{12}^{i} + 12 (S_{2}v_{12}v_{2})v_{2}^{i} \;,\\
    \gamma^{i}_{2,0} &= S_{2}^{ij}n_{12}^{j} 
\left[ -\frac{1815}{8} (n_{12}v_{12})^3 - 
51 (n_{12}v_{12})^2 (n_{12}v_{2}) + 54 (n_{12}v_{12}) (n_{12}v_{2})^2 - 
6 (n_{12}v_{2})^3  \right. \nonumber \\
& \qquad \qquad \quad \left. + 
\frac{801}{8} (n_{12}v_{12}) v_{12}^{2} + 12 (n_{12}v_{2}) v_{12}^{2} + 
\frac{31}{2} (n_{12}v_{12}) (v_{12}v_{2}) - 
\frac{39}{2} (n_{12}v_{2}) (v_{12}v_{2})  \right. \nonumber \\
& \qquad \qquad \quad \left.  + \frac{31}{4} (n_{12}v_{12}) v_{2}^{2} -  
(n_{12}v_{2}) v_{2}^{2} \right] \nonumber \\
& + S_{2}^{ij}v_{12}^{j} \left[ \frac{1087}{8} (n_{12}v_{12})^2 + 
56 (n_{12}v_{12}) (n_{12}v_{2}) - \frac{59}{2} (n_{12}v_{2})^2 - 
\frac{269}{8} v_{12}^{2} \right. \nonumber \\
& \qquad \qquad \; \left.  - \frac{55}{2} (v_{12}v_{2}) - 
\frac{23}{4} v_{2}^{2} \right] \nonumber \\
& + (S_{2}n_{12}v_{12})n_{12}^{i} 
\left[ -\frac{1797}{8} (n_{12}v_{12})^2 - 
54 (n_{12}v_{12}) (n_{12}v_{2}) + 81 (n_{12}v_{2})^2 + 
\frac{323}{8} v_{12}^{2}   \right. \nonumber \\
& \qquad \qquad \qquad \qquad \left.  + 
\frac{93}{2} (v_{12}v_{2}) + \frac{45}{4} v_{2}^{2} \right] \nonumber \\
& + (S_{2}n_{12}v_{12})v_{12}^{i} \left[ 67 (n_{12}v_{12}) + 
\frac{35}{2} (n_{12}v_{2}) \right] - 
24 (n_{12}v_{12}) (S_{2}n_{12}v_{12})v_{2}^{i} \nonumber \\
& + (S_{2}n_{12}v_{2})v_{12}^{i} \left[ -\frac{31}{2} (n_{12}v_{12}) + 
2 (n_{12}v_{2}) \right] + (S_{2}n_{12}v_{2})v_{2}^{i} 
\left[ -\frac{31}{2} (n_{12}v_{12}) + 2 (n_{12}v_{2}) \right] \nonumber \\
& + \frac{23}{2} (S_{2}v_{12}v_{2})v_{12}^{i} + 
\frac{23}{2} (S_{2}v_{12}v_{2})v_{2}^{i} \;,\\
    \gamma^{i}_{0,3} &= \frac{15}{2} (n_{12}v_{12}) S_{1}^{ij}n_{12}^{j} - 
\frac{15}{2} S_{1}^{ij}v_{12}^{j} + 15 (S_{1}n_{12}v_{12})n_{12}^{i} \;,\\
    \gamma^{i}_{1,2} &= 
\frac{227}{8} (n_{12}v_{12}) S_{1}^{ij}n_{12}^{j} - 
\frac{309}{8} S_{1}^{ij}v_{12}^{j} + 
\frac{691}{8} (S_{1}n_{12}v_{12})n_{12}^{i} \nonumber \\
& + 32 (n_{12}v_{12}) S_{2}^{ij}n_{12}^{j} - 
24 S_{2}^{ij}v_{12}^{j} + 42 (S_{2}n_{12}v_{12})n_{12}^{i} \;,\\
    \gamma^{i}_{2,1} &= \frac{79}{2} (n_{12}v_{12}) S_{1}^{ij}n_{12}^{j} - 
\frac{63}{2} S_{1}^{ij}v_{12}^{j} + 
58 (S_{1}n_{12}v_{12})n_{12}^{i} \nonumber \\
& + S_{2}^{ij}n_{12}^{j} \left[ \frac{251}{4} (n_{12}v_{12}) - 
14 (n_{12}v_{2}) \right] - 61 S_{2}^{ij}v_{12}^{j} + 
\frac{257}{2} (S_{2}n_{12}v_{12})n_{12}^{i} \;,\\
    \gamma^{i}_{3,0} &= S_{2}^{ij}n_{12}^{j} 
\left[ -\frac{73}{8} (n_{12}v_{12}) - 14 (n_{12}v_{2}) \right] - 
\frac{119}{8} S_{2}^{ij}v_{12}^{j} + 
\frac{343}{8} (S_{2}n_{12}v_{12})n_{12}^{i}\;.
\end{align}
\end{subequations}}
%


\section{Checks of the results}
\label{secChecks}

\subsection{Conserved Energy}
\label{subsecEnergy}

An important feature of all the spin-orbit contributions we have computed, is
that they are associated with the \textit{conservative} part of the dynamics,
\textit{i.e.} obtained when neglecting the radiation-reaction dissipative
terms associated with gravitational radiation. Therefore these contributions
should allow for the existence of a set of conserved quantities, namely
energy, angular momentum, linear momentum, and center-of-mass integral.
Here we have checked that the new terms computed in the equations of motion in
Sec.~\ref{subsecAcc} admit corresponding contributions in the conserved
energy, which take the following structure:
\begin{align}\label{Estruct}
E &=
E_\mathrm{N}+\frac{1}{c^2}E_\mathrm{1PN}+\frac{1}{c^3}
\mathop{E}_{S}{}_{\!\mathrm{1.5PN}}
+\frac{1}{c^4}\left[E_\mathrm{2PN}+
\mathop{E}_{SS}{}_{\!\mathrm{2PN}}\right]
+\frac{1}{c^5} \mathop{E}_{S}{}_{\!\mathrm{2.5PN}} \nonumber\\
& \qquad + \frac{1}{c^6}\left[E_\mathrm{3PN}+
\mathop{E}_{SS}{}_{\!\mathrm{3PN}}\right] + 
\frac{1}{c^7} \mathop{E}_{S}{}_{\!\mathrm{3.5PN}} + 
\mathcal{O}\left(\frac{1}{c^8}\right) \,.
\end{align}
The leading order spin-orbit contribution to the energy reads
\begin{equation}
	\mathop{E}_{S}{}_{\!\mathrm{1.5PN}} = \frac{G m_{2}}{c^{3} r_{12}^{2}} 
(S_{1}n_{12}v_{1}) + 1\leftrightarrow2 \;,
\end{equation}
while the 2.5PN result (1PN relative) was given in Paper~I, and reads, once
translated into the spin tensor variable $S^{ij}$:
\begin{equation}
	\mathop{E}_{S}{}_{\!\mathrm{2.5PN}} = \frac{G}{r_{12}^{2}} 
\epsilon_{0,1} m_{2} + \frac{G^{2}}{r_{12}^{3}} \left[ \epsilon_{0,2}
  m_{2}^{2} + \epsilon_{1,1} m_{1}m_{2} \right] + 1\leftrightarrow2 \;,
\end{equation}
where
\begin{subequations}
\begin{align}
    \epsilon_{0,1} &= (S_{1}n_{12}v_{12}) 
\left[ 3 (n_{12}v_{12}) (n_{12}v_{2}) + 
\frac{9}{2} (n_{12}v_{2})^2 + (v_{12}v_{2}) \right] \nonumber \\
& + (S_{1}n_{12}v_{2}) \left[ -3 (n_{12}v_{12})^2 - 
6 (n_{12}v_{12}) (n_{12}v_{2}) - \frac{3}{2} (n_{12}v_{2})^2 + 
v_{12}^{2} \right] \nonumber \\
& + (S_{1}v_{12}v_{2}) \left[ (n_{12}v_{12}) + 3 (n_{12}v_{2}) \right] \;,\\
    \epsilon_{0,2} &= 2 (S_{1}n_{12}v_{12}) - (S_{1}n_{12}v_{2}) \;,\\
    \epsilon_{1,1} &= -2 (S_{1}n_{12}v_{12}) - 2 (S_{1}n_{12}v_{2}) \;.
\end{align}
\end{subequations}
The result for the 3.5PN contribution to the conserved energy has
been obtained by the method of unknown coefficients and successfully
determines the following unique result:
\begin{align}
	\mathop{E}_{S}{}_{\!\mathrm{3.5PN}} &= 
\frac{G}{r_{12}^{2}} \eta_{0,1} m_{2} + \frac{G^{2}}{r_{12}^{3}} 
\left[ \eta_{0,2} m_{2}^{2} + \eta_{1,1} m_{1}m_{2} \right] \nonumber \\
	& \quad + \frac{G^{3}}{r_{12}^{4}} \left[ \eta_{0,3} m_{2}^{3} + 
\eta_{1,2} m_{1}m_{2}^{2}  + \eta_{2,1} m_{1}^{2}m_{2} \right] + 
1\leftrightarrow2 \;,
\end{align}
where
\allowdisplaybreaks{
\begin{subequations}
\begin{align}
    \eta_{0,1} &= (S_{1}n_{12}v_{12}) 
\left[ -\frac{15}{4} (n_{12}v_{12})^3 (n_{12}v_{2}) - 
15 (n_{12}v_{12})^2 (n_{12}v_{2})^2 - 
\frac{45}{2} (n_{12}v_{12}) (n_{12}v_{2})^3 \right. \nonumber \\
& \qquad \qquad \qquad \left. - \frac{105}{8} (n_{12}v_{2})^4 + 
\frac{9}{4} (n_{12}v_{12}) (n_{12}v_{2}) v_{12}^{2} + 
3 (n_{12}v_{2})^2 v_{12}^{2} - 
\frac{3}{4} (n_{12}v_{12})^2 (v_{12}v_{2}) \right. \nonumber \\
& \qquad \qquad \qquad \left. + 
3 (n_{12}v_{12}) (n_{12}v_{2}) (v_{12}v_{2}) + 
\frac{9}{2} (n_{12}v_{2})^2 (v_{12}v_{2}) + 
\frac{1}{4} v_{12}^{2} (v_{12}v_{2}) + (v_{12}v_{2})^2  \right. \nonumber \\
& \qquad \qquad \qquad \left. + 
3 (n_{12}v_{12})^2 v_{2}^{2} + 12 (n_{12}v_{12}) (n_{12}v_{2}) v_{2}^{2} + 
\frac{21}{2} (n_{12}v_{2})^2 v_{2}^{2} -  v_{12}^{2} v_{2}^{2} + 
2 (v_{12}v_{2}) v_{2}^{2} \right] \nonumber \\
& + (S_{1}n_{12}v_{2}) \left[ \frac{15}{4} (n_{12}v_{12})^4 + 
15 (n_{12}v_{12})^3 (n_{12}v_{2}) + 
\frac{45}{2} (n_{12}v_{12})^2 (n_{12}v_{2})^2  \right. \nonumber \\
& \qquad \qquad \qquad \left. + 15 (n_{12}v_{12}) (n_{12}v_{2})^3 + 
\frac{15}{8} (n_{12}v_{2})^4 - \frac{9}{2} (n_{12}v_{12})^2 v_{12}^{2} - 
9 (n_{12}v_{12}) (n_{12}v_{2}) v_{12}^{2}  \right. \nonumber \\
& \qquad \qquad \qquad \left. - \frac{9}{2} (n_{12}v_{2})^2 v_{12}^{2} + 
\frac{3}{4} v_{12}^{4} - 6 (n_{12}v_{12})^2 (v_{12}v_{2}) - 
12 (n_{12}v_{12}) (n_{12}v_{2}) (v_{12}v_{2})  \right. \nonumber \\
& \qquad \qquad \qquad \left. - 6 (n_{12}v_{2})^2 (v_{12}v_{2}) + 
2 v_{12}^{2} (v_{12}v_{2}) + (v_{12}v_{2})^2 - 
3 (n_{12}v_{12})^2 v_{2}^{2}  \right. \nonumber \\
& \qquad \qquad \qquad \left. - 6 (n_{12}v_{12}) (n_{12}v_{2}) v_{2}^{2} - 
\frac{3}{2} (n_{12}v_{2})^2 v_{2}^{2} + v_{12}^{2} v_{2}^{2} \right] \nonumber \\
& + (S_{1}v_{12}v_{2}) \left[ -\frac{3}{4} (n_{12}v_{12})^3 - 
3 (n_{12}v_{12})^2 (n_{12}v_{2}) - 
\frac{9}{2} (n_{12}v_{12}) (n_{12}v_{2})^2 - 
\frac{9}{2} (n_{12}v_{2})^3  \right. \nonumber \\
& \qquad \qquad \qquad \left. + \frac{3}{4} (n_{12}v_{12}) v_{12}^{2} + 
(n_{12}v_{2}) v_{12}^{2} + (n_{12}v_{12}) (v_{12}v_{2}) + 
(n_{12}v_{2}) (v_{12}v_{2})  \right. \nonumber \\
& \qquad \qquad \qquad \left. + (n_{12}v_{12}) v_{2}^{2} + 
3 (n_{12}v_{2}) v_{2}^{2} \right]  \;,\\
    \eta_{0,2} &= (S_{1}n_{12}v_{12}) \left[ 23 (n_{12}v_{12}) (n_{12}v_{2}) - 
\frac{53}{2} (n_{12}v_{2})^2 - \frac{109}{8} (v_{12}v_{2}) -  
v_{2}^{2} \right] \nonumber \\
& + (S_{1}n_{12}v_{2}) \left[ -\frac{53}{2} (n_{12}v_{12})^2 + 
\frac{35}{2} (n_{12}v_{12}) (n_{12}v_{2}) + 3 (n_{12}v_{2})^2 + 
\frac{65}{8} v_{12}^{2} \right] \nonumber \\
& + (S_{1}v_{12}v_{2}) \left[ \frac{87}{8} (n_{12}v_{12}) - 
\frac{29}{2} (n_{12}v_{2}) \right] \;,\\
    \eta_{1,1} &= (S_{1}n_{12}v_{12}) 
\left[ -\frac{105}{4} (n_{12}v_{12})^2 - 
\frac{141}{2} (n_{12}v_{12}) (n_{12}v_{2}) - 22 (n_{12}v_{2})^2 + 
\frac{7}{2} v_{12}^{2} + 18 (v_{12}v_{2}) \right] \nonumber \\
& + (S_{1}n_{12}v_{2}) \left[ \frac{161}{4} (n_{12}v_{12})^2 + 
22 (n_{12}v_{12}) (n_{12}v_{2}) + 4 (n_{12}v_{2})^2 - 
\frac{13}{2} v_{12}^{2} \right] \nonumber \\
& + (S_{1}v_{12}v_{2}) \left[ -\frac{123}{4} (n_{12}v_{12}) - 
14 (n_{12}v_{2}) \right]  \;,\\
    \eta_{0,3} &= 2 (S_{1}n_{12}v_{12}) - \frac{5}{4} (S_{1}n_{12}v_{2})  \;,\\
    \eta_{1,2} &= \frac{41}{4} (S_{1}n_{12}v_{12}) + 
\frac{27}{4} (S_{1}n_{12}v_{2})  \;,\\
    \eta_{2,1} &= \frac{15}{4} (S_{1}n_{12}v_{12}) + 
\frac{15}{4} (S_{1}n_{12}v_{2})\;. 
\end{align}
\end{subequations}}\noindent
We leave for future work the study of the other conserved quantities at the
same order, namely the total angular momentum, linear momentum, and
center-of-mass integral.

\subsection{Lorentz invariance}
\label{subsecLorentz}

Since we are working with the harmonic gauge condition
which is manifestly Lorentz invariant, the global Lorentz invariance must be
preserved by our calculations and must be manifest on our final equations of
motion.\footnote{The global Lorentz invariance is the one associated with the background flat space-time, which approaches the asymptotically Minkowskian space-time far away from the compact-support matter distribution.} To check this,
we follow mostly the presentation of Ref.~\cite{BFregM} and of the Appendix~A
of Paper~I, with the difference that we are using a different spin variable,
which will simplify this calculation.

Let us consider two different frames $(\calF)$ and $(\calF')$, the latter
being related to the former by a boost of velocity $\bfV$. The coordinates
of a given space-time event $P$ are $x^{\mu}$ in the original frame $(\calF)$
and $x'^{\mu}$ in the boosted frame $(\calF')$, both being related by the
Lorentz transformation $x'^{\mu}=\Lambda^{\mu}_{\ph{\mu}\nu}x^{\nu}$, with:
\begin{subequations}\label{boost}
\begin{align}
	\Lambda^{0}_{\ph{0}0} &= \gamma \;, \\
	\Lambda^{i}_{\ph{i}0} &= \Lambda^{0}_{\ph{0}i}= - \gamma \frac{V^{i}}{c} \;, \\
	\Lambda^{i}_{\ph{i}j} &= \delta^{i}_{\ph{i}j} + 
    \frac{\gamma^{2}}{\gamma+1} \frac{V^{i}V_{j}}{c^{2}} \;,
\end{align}
\end{subequations}
with $\gamma=(1-V^{2}/c^{2})^{-1/2}$ the Lorentz factor. We denote the
trajectories of the two bodies in the frame $(\calF)$ by
$y^{\mu}_{1}=(ct,\bfy_{1})$ and $y^{\mu}_{2}=(ct,\bfy_{2})$, and by
$y'^{\mu}_{1}=(ct',\bfy'_{1})$ and $y'^{\mu}_{2}=(ct',\bfy'_{2})$ in the
boosted frame $(\calF')$. The point is that we cannot compare them directly,
because the simultaneity surfaces are different in the two different frames.
To give a sense to simultaneity between the two frames, it is convenient to
define an auxiliary event $\Omega(ct,\bfx)$, which can be choosen at will ---
it could be for instance the point of coordinates $(ct, \bfx=0)$ 
in $(\calF)$, and
whose coordinates in $(\calF')$ will be $(ct',\bfx')$. We define the two
events $P_{1}(ct,\bfy_{1}(t))$ and $P_{2}(ct,\bfy_{2}(t))$ on the two
worldines, simultaneous to $\Omega$ in the frame $(\calF)$, and similarly two
events $Q_{1}(ct',\bfy'_{1}(t'))$ and $Q_{2}(ct',\bfy'_{2}(t'))$, simultaneous
to $\Omega$ in the frame $(\calF')$. $\Omega$ plays the role of an observer,
for which the equations of motion evaluated on the simultaneity surfaces
$(\Omega,P_{1},P_{2})$ and $(\Omega,Q_{1},Q_{2})$ must take the same
fonctional form, as explained below. Next, we define the times $\tau_{1}$ and
$\tau_{2}$, such that the coordinates of $Q_{1}$ and $Q_{2}$ in $(\calF)$ are
$(c\tau_{1},\bfy_{1}(\tau_{1}))$ and $(c\tau_{2},\bfy_{2}(\tau_{2}))$. We
have, by construction, ${y'_{1}}^{\mu}(t') = \Lambda^{\mu}_{\ph{\mu}\nu}
y_{1}^{\nu}(\tau_{1})$
and similarly for 2. The link between ${\bfy'_{1}}(t')$ and ${\bfy_{1}}(t)$ is
provided in Ref.~\cite{BFregM}. One obtains successively
\begin{subequations}\label{var1prime}
\begin{align}
	\mathbf{y}'_{1}(t') &= \mathbf{y}_1 - \gamma \mathbf{V} 
\left(t - \frac{1}{c^2} \frac{\gamma}{\gamma + 1} (Vx)\right) + 
\sum_{n=1}^{+\infty} \frac{(-)^n}{c^{2n} n!} \partial_t^{n-1} 
\left[ (Vr_1)^n \left(\mathbf{v}_{1} - 
\frac{\gamma}{\gamma + 1} \mathbf{V} \right) \right]\;, \label{y1prime} \\
	\mathbf{v}'_{1}(t') &= \frac{\mathbf{v}_1}{\gamma} - 
\mathbf{V} + \frac{1}{\gamma} \sum_{n=1}^{+\infty} \frac{(-)^n}{c^{2n} n!} 
\partial_t^n \left[(Vr_1)^n \left(\mathbf{v}_1 - 
\frac{\gamma}{\gamma + 1} \mathbf{V}\right) \right] \;, \label{v1prime} \\
	\mathbf{a}'_{1}(t') &= \frac{1}{\gamma^2} 
\left\{ \mathbf{a}_1 + 
\sum_{n=1}^{+\infty} \frac{(-)^{n}}{c^{2n} n!} 
\partial_t^{n+1} \left[(Vr_1)^n \left(\mathbf{v}_1 - 
\frac{\gamma}{\gamma + 1} \mathbf{V}\right) \right] \right\} 
\label{a1prime} \;.
\end{align}
\end{subequations}
where the right-hand sides are evaluated at $t$, and $(Vr_{1})$ means
$\bfV\cdot(\bfx-\bfy_{1})$. The expressions for the velocity and acceleration
are obtained by taking the time derivative according to $\partial'_t =
\gamma \partial_t + \gamma V^i \partial_i$. Similar expressions hold for 2,
and transformations for quantities such as $r_{12}$ or $n_{12}^{i}$ are
deduced in a perturbative sense from the first of these formulae.

We turn now to the transformation rules for the spin tensor. As explained in
Paper~I, we have
for any function $f(t)$:
\begin{equation}
	f(\tau_1) = f(t) + \sum_{n = 1}^{+ \infty} \frac{(-)^n}{c^{2n} n!} 
\partial_t^{n - 1} \left[ \frac{\ud f}{\ud t} (Vr_1)^n \right] \; ,
\end{equation}
which can be applied to the components $S_{1}^{0i}$ and $S_{1}^{ij}$ of the
spin tensor. Now, since these are the components of a contravariant tensor, we
have
\begin{subequations}\label{S1primelorentz}
\begin{align}
	{S'_{1}}^{0i}(t') &= \Lambda^{0}_{\ph{0}\mu} \Lambda^{i}_{\ph{i}\nu} 
S_{1}^{\mu\nu}(\tau_{1}) \;,  \\
	{S'_{1}}^{ij}(t') &= \Lambda^{i}_{\ph{i}\mu} \Lambda^{j}_{\ph{j}\nu} 
S_{1}^{\mu\nu}(\tau_{1}) \;.
\end{align}
\end{subequations}
Combining these expressions, we arrive at the desired transformation rule for
$S_{1}^{ij}$:
\begin{align} \label{S1prime}
	{S'_{1}}^{ij}(t') &= S_{1}^{ij} - 
\frac{2\gamma^{2}}{\gamma+1}\frac{V^{k}}{c^{2}}S_{1}^{k[i}V^{j]} +
\frac{2\gamma}{c}S_{1}^{0[i}V^{j]} \nonumber \\
	& + \sum_{n=1}^{+\infty} \frac{(-)^{n}}{c^{2n} n!} 
\partial_t^{n-1} \left[(Vr_1)^n \left( \frac{\ud S_{1}^{ij}}{\ud t} - 
\frac{2\gamma^{2}}{\gamma+1}\frac{V^{k}}{c^{2}}
\frac{\ud S_{1}^{k[i}}{\ud t}V^{j]} + 
\frac{2\gamma}{c}\frac{\ud S_{1}^{0[i}}{\ud t}V^{j]} \right) \right] \;,
\end{align}
where $S_{1}^{0i}$ can be further eliminated using the spin supplementary
condition \eqref{S0itoSij}. This transformation is somewhat simpler than
(A13)-(A15) in Appendix~A of Paper~I
because the transformation for the spin vector $S_\text{FBB}^{i}$ is not as
simple as in Eqs.~\eqref{S1primelorentz}.

Lorentz invariance means that the acceleration in the boosted frame, obtained
by transforming all variables according to Eqs.~\eqref{var1prime} and
\eqref{S1prime}, when systematically truncated at the requested PN order, must
take the same \textit{functional} form as in the original non-boosted frame.
We have verified that our result for the acceleration \eqref{a1iS7} passes
this test. Note however that the test leaves a lot of freedom in the
acceleration. In particular,
one might add at the highest level $1/c^{7}$ in the acceleration any quantity
depending only on the relative velocity $v_{12}^i=v_{1}^i-v_{2}^i$ and still
pass this test.

Finally let us remark that the pure Hadamard Schwartz regularization which has
been used for all the terms in this computation, finally yields equations of
motion which are manifestly Lorentz invariant. This is another indication that
there is no need for using a more sophisticated regularization such as
dimensional regularization in this problem.

\subsection{Test-mass limit}
\label{subsecTestmass}

Another important check of our result is to take the test mass limit and show
that we recover the equations of motion of a test particle orbiting in a black
hole background. More precisely, we can check that:
\begin{enumerate}
\item In the limit where one of the two bodies (say body $1$) is a test
  particle with $m_1\rightarrow 0$ and is spinless ($S_1^{ij}=0$), its
  acceleration \eqref{a1struct} reduces to that of a spinless test particle
  orbiting around a Kerr black hole;
\item In the limit where body $1$ is a test particle with spin (\textit{i.e.}
  $m_1\rightarrow0$ with constant ratio $S_1^{ij}/m_1$), Eq.~\eqref{a1struct}
  reduces to the equations of motion of a massless spinning particle orbiting
  around a Schwarzschild black hole.
\end{enumerate}
Note that in case 2.
we need to work with a Schwarzschild black hole because we restricted
ourselves to spin-orbit effects. In this section, it will be more convenient
to rewrite the equation of motion
\eqref{eomstructure}--\eqref{eomstructuredef} so as to make apparent the
coordinate acceleration:
\begin{equation}\label{eomcoordinate}
	\frac{\ud v^{\mu}}{\ud t} = \frac{\ud^{2} x^{\mu}}{\ud t^{2}} = 
v^{\nu}v^{\rho}\left( \frac{v^{\mu}}{c}\Gamma^{0}_{\ph{0}\nu\rho} - 
\Gamma^{\mu}_{\ph{\mu}\nu\rho} \right) + \frac{1}{u^{0}} 
\left( \calF^{\mu} - \frac{v^{\mu}}{c}\calF^{0} \right) \; ,
\end{equation}
where we recall that $\calF^{\mu} = - \frac{1}{2 m c} 
R^{\mu}_{\ph{\mu}\nu\rho\sigma}v^{\nu}S^{\rho\sigma}$ and 
$u^{0}=1/\sqrt{-g_{\rho\sigma}v^{\rho}v^{\sigma}/c^{2}}$.

\subsubsection{Spinless test particle: equivalence with Kerr geodesics}

In the limit where $m_1\rightarrow0$ and $S_1^{ij}\rightarrow0$, then
$v_2^{i}=0$ and $S_2^{ij}=\text{const}$ is a trivial solution of the
equations of motion and the (spin part of the) acceleration of body $1$ given
by Eq.~\eqref{a1iS7}, simply reduces to
\begin{eqnarray}
\label{PNtestmasslimitnospin0}
\nonumber (a_1^i)_S&=& 
 \frac{G}{c^3 r_{12}^3} \left(6 (n_{12}v_1) A_2^i - 4 B_2^i + 
6 (S_2n_{12}v_1) n_{12}^i\right) - 
6 \frac{G}{c^5 r_{12}^3}(n_{12}v_1) (S_2n_{12}v_1) v_1^{i}\\ \nonumber
 && + \frac{G^2}{c^5 r_{12}^4} \left(
 -16 m_2 (n_{12}v_1) A_2^i
 - 20 m_2 (S_2n_{12}v_1) n_{12}^i
 +12 m_2 B_2^i\right)\\
 && + \frac{G^3}{c^7 r_{12}^5} \left(
 32 m_2^2 (n_{12}v_1) A_2^i
 +42 m_2^2 (S_2n_{12}v_1) n_{12}^i
 -24 m_2^2 B_2^i \right)\;,
\end{eqnarray}
where we have defined $A_2^i=n_{12}^j S_2^{ij}$ and $B_2^i=v_1^j S_2^{ij}$.
For simplicity, we choose the origin of our coordinate system at the location
of the central body $2$ and suppose without loss of generality that the
spin of body $2$ points along the $z$ axis: defining
$S^{i}=\varepsilon^{ijk}S_2^{jk}$, we impose $S^x=S^y=0$ and $S^z=m_2 \,a$.

Given the symmetry of the problem, it is of course more convenient to work
with the spherical coordinates associated with our harmonic coordinates
and with the associated coordinate basis that we denote
$(\partial_r,\partial_\theta,\partial_\phi)$. In practice, we will show that
our PN result recovers the dynamics of a test-mass in a Kerr background by
comparing the explicit expressions for the spin parts of the quantities
$\ddot{r}$, $\ddot{\theta}$ and $\ddot{\phi}$ obtained using
Eq.~\eqref{PNtestmasslimitnospin0} and the geodesic equation in Kerr. In terms
of $r$, $\theta$ and $\phi$, we have the following set of scalar quantities
$r_{12}=r$, $(n_{12}v_1)=\dot{r}$, $(S_2n_{12}v_1)=-m_2 a r\dot \phi \sin ^2
\theta $, and vector components $v_1^r=\dot{r}$, $v_1^\theta=\dot{\theta}$,
$v_1^\phi=\dot{\phi}$, $n_{12}^r=1$, $A_2^\phi=-m_2 a/r$, $B_2^r=-m_2 a r
\dot{\phi}\sin^2\theta $, $B_2^\theta=-m_2 a \dot{\phi}\cos\theta\sin\theta$
and $B_2^\phi=m_2 a(\dot{\theta}\cot{\theta}+\dot{r}/r)$, all the other ones
being zero. The components of the Cartesian acceleration are
$a^r=\ddot{r}-r\dot{\theta}^2-r\dot{\phi}^2\sin^2\theta$,
$a^\theta=\ddot{\theta}+2\dot{r}\dot{\theta}/r-\dot{\phi}^2\sin\theta\cos\theta$
and $a^\phi=\ddot\phi+2\dot{r}\dot\phi \cot\theta/r
+2\dot\theta\dot\phi\cot\theta$. This yields
\begin{subequations}\label{Kerreqspp}
\begin{align}
\label{Kerrrpp}
(\ddot{r})_S&=2\frac{Gm_2 a}{c^3 r^2}\dot\phi \sin^2\theta
-6\frac{Gm_2 a}{c^5 r^2}\dot\phi\, \dot{r}^2 \sin^2\theta
-8\frac{G^2m_2^2 a}{c^5 r^3}\dot\phi \sin^2\theta
+18\frac{G^3m_2^3 a}{c^7 r^4}\dot\phi \sin^2\theta\;,\\
\nonumber\label{Kerrthetapp}
(\ddot{\theta})_S&=
-4\frac{Gm_2 a}{c^3 r^3}\dot\phi \cos\theta\sin\theta
-6\frac{Gm_2 a}{c^5 r^2}\dot{r} \, \dot\theta \, \dot\phi  \sin^2\theta
+12\frac{G^2m_2^2 a}{c^5 r^4}\dot\phi \cos\theta\sin\theta\\
&\qquad -24\frac{G^3m_2^3 a}{c^7 r^5}\dot\phi\cos\theta \sin\theta\;,\\
\nonumber\label{Kerrphipp}
(\ddot{\phi})_S&=
\frac{Gm_2 a}{c^3 r^4}(4r\dot\theta \cot\theta-2\dot{r})
-6\frac{Gm_2 a}{c^5 r^2}\dot{r}\dot{\phi}^2\sin^2\theta
+\frac{G^2m_2^2 a}{c^5 r^5}(4\dot{r}-12 r \dot\theta \cot\theta)\\
&\qquad+24\frac{G^3m_2^3 a}{c^7 r^5}\dot\theta \cot\theta\;.
\end{align}
\end{subequations}
We would like to compare this with the equations of motion of a test particle
in the background of a Kerr black hole of mass $m_2$ and spin parameter $a$.
Since our result is written in the harmonic gauge, we also need to work with
the Kerr metric in harmonic coordinates rather than in the usual
Boyer-Lindquist (BL) ones. A particular set of spatial harmonic coordinates
($x^1$, $x^2$ and $x^3$) constructed from the BL grid was obtained in
Ref.~\cite{CS97} and reads
\begin{subequations}\label{Kerrharmcoord}
\begin{align}
\label{Kerrharmcoords12}
x^1+\ui x^2&= \left(r_\text{BL} - m_2+ \ui a\right) 
\sin \theta_\text{BL} \exp \left(\ui \left[\phi_\text{BL}+
\frac{a}{r_+-r_-}\ln\left| 
\frac{r_\text{BL}-r_+}{r_\text{BL}-r_-}\right|\,\right]\right)\;,\\
\label{Kerrharmcoord3}
x^3&=\left(r_\text{BL}-m_2\right) \cos\theta_\text{BL}\;,
\end{align}
\end{subequations}
with $r_\pm=(m_2\pm \sqrt{m_2^2-a^2})$. For the time coordinate, we can simply
choose $t=t_\text{BL}$. Here again, we will use the spherical coordinates
associated to $x^1$, $x^2$ and $x^3$. The line element, expanded to linear
order in the spin $a$, reads
\begin{eqnarray}
\label{linelement}
\ud s^2&=&
\nonumber-\frac{r-m_2}{r+m_2}\ud t^2
+\frac{r+m_2}{r-m_2}\ud r^2
+\left(r+m_2\right)^2(\ud\theta^2+\sin^2\theta\ \ud\phi^2)\\
&&-2\frac{m_2^2\, a}{r^2}\frac{r+m_2}{r-m_2}\sin^2\theta \ud r \ud\phi
-4 m_2\, a\frac{\sin^2\theta}{r+m_2}\;\ud t\ud\phi
+\mathcal{O}(a^2)\;.
\end{eqnarray}
Note that in order to avoid heavy notations, we have set $G=c=1$ in these non
PN-expanded equations \eqref{Kerrharmcoord}--\eqref{linelement}. 
Using \eqref{eomcoordinate}, in which we set $\calF^{\mu}$ to zero, for the
coordinates $x'^\mu=(t,r,\theta,\phi)$, and developing at linear order in $a$
and at 3.5PN order, we obtain expressions for $\ddot{r}$, $\ddot{\theta}$ and
$\ddot{\phi}$ that reduce to the results of Eqs.~\eqref{Kerreqspp}, but with
the addition of the following extra pure gauge terms at 3PN order:
\begin{subequations}
\begin{align}
\left(\delta  \ddot{r}\right) _S&=-
2\frac{G^2m_2^2 a}{c^6 r^3}\dot{r}\dot\phi \sin^2\theta \;,\\
\left(\delta  \ddot{\theta}\right)_S&=-
2\frac{G^2m_2^2 a}{c^6 r^4}\dot{r}\dot\phi \cos\theta\sin\theta \;,\\
\left(\delta  \ddot{\phi}\right)_S&=
-\frac{G^2m_2^2 a}{c^6 r^5}(2\dot{r}^2-
2r \dot{r}\dot{\theta}\cot\theta-
r^2(\dot{\theta}^2+\dot{\phi}^2\sin^2\theta))
-\frac{G^3m_2^3 a}{c^6 r^6}\;.
\end{align}
\end{subequations}
The presence of these terms is simply a consequence of the fact that the
coordinates \eqref{Kerrharmcoord} are not exactly the same as the ones we used
in our PN calculations: the harmonicity condition does not completely fix the
gauge and we still have the freedom to perform a coordinate change
$x^\mu\rightarrow x^\mu+\xi^\mu$ as long as
$\partial_\rho(\sqrt{-g}g^{\rho\sigma}\partial_\sigma \xi^\mu)=0$ without
violating the harmonicity condition. More precisely, the presence of these
terms can be traced back to the metric element $g_{r\phi}$ which can easily be
put to zero (modulo higher order PN corrections) by the coordinate shift
\begin{equation}
\phi\rightarrow \phi+  \frac{2G^2m_2^2a}{3c^6 r^3}\;.
\end{equation}
In terms of our harmonic coordinates, this translates into a shift
$\xi^1=-\frac{2G^2m_2^2a}{3c^6}\frac{x^2}{r^3}$ and
$\xi^2=\frac{2G^2m_2^2a}{3c^6}\frac{x^1}{r^3}$ which can easily be seen to
satisfy $\partial_\rho(\sqrt{-g}g^{\rho\sigma}\partial_\sigma \xi^\mu)=0$
modulo higher than 3.5PN corrections: since $\xi^\mu=\mathcal{O}(1/c^6)$ (and
it does not depend on time), this equation simply becomes $\Delta\xi^i=0$.

\subsubsection{Spinning test particle: equivalence with Papapetrou motion 
in Schwarzschild}

We now consider the limit $m_1\rightarrow0$ while $S_1^{ij}/m_1$ is kept
constant and we set $S_2^{ij}=0$ since we only want to study linear effects in
the spins (the spin-orbit terms involving $S_2$ are precisely the ones that
have been studied in the previous subsection). Here again, $v_2^i=0$ is a
solution of the equations of motion and the acceleration of body $1$ reduces
to
\begin{eqnarray}
\label{PNtestmasslimitnospin}
\nonumber (a_1^i)_S&=& 
 \frac{G}{c^3 r_{12}^3}\frac{m_2}{m_1} \left(3 (n_{12}v_1) A_1^i + 
6 (S_1n_{12}v_1) n_{12}^i - 3 B_1^i\right)\\ 
\nonumber && +\frac{G}{c^5 r_{12}^3}\frac{m_2}{m_1} 
\left(-\frac{3}{2}(n_{12}v_1)v_1^2A_1^i -
3v_1^2 (S_1n_{12}v_1) n_{12}^i+\frac{3}{2}v_1^2B_1^i-
3(n_{12}v_1) (S_1n_{12}v_1)v_1^i\right)\\
\nonumber  && +\frac{G^2}{c^5 r_{12}^4}\frac{m_2^2}{m_1} 
\left(-6 (n_{12}v_1) A_1^i -12 (S_1n_{12}v_1) n_{12}^i +6 B_1^i\right)\\  
\nonumber &&  +\frac{G}{c^7 r_{12}^3}\frac{m_2}{m_1} v_1^2
\left(-\frac{3}{8}(n_{12}v_1)v_1^2A_1^i -
\frac{3}{4}v_1^2 (S_1n_{12}v_1) n_{12}^i+
\frac{3}{8}v_1^2B_1^i+\frac{3}{2}(n_{12}v_1) (S_1n_{12}v_1)v_1^i\right)\\ 
\nonumber &&  +\frac{G^2}{c^7 r_{12}^4}\frac{m_2^2}{m_1}
\left(-3(n_{12}v_1)v_1^2A_1^i -6v_1^2 (S_1n_{12}v_1) n_{12}^i+
3v_1^2B_1^i -6(n_{12}v_1) (S_1n_{12}v_1)v_1^i\right)\\
&&  +\frac{G^3}{c^7 r_{12}^5}\frac{m_2^3}{m_1}
\left(\frac{15}{2}(n_{12}v_1)A_1^i +15 (S_1n_{12}v_1) n_{12}^i-
\frac{15}{2}B_1^i\right)\;,
\end{eqnarray}
with $A_1^i=n_{12}^j S_1^{ij}$ and $B_1^i=v_1^j S_1^{ij}$. Choosing once again
the origin of the coordinate system at the position of body $2$ and moving to
spherical coordinates, we obtain $A_1^r=0$, $A_1^\theta=-S_1^{r\theta}$,
$A_1^\phi=-S_1^{r\phi}$, $B_1^r=r^2\dot\theta
S_1^{r\theta}+r^2\dot\phi\sin^2\theta S_1^{r\phi}$,
$B_1^\theta=-\dot{r}S_1^{r\theta}+r^2\dot\phi\sin^2\theta S_1^{\theta\phi}$,
$B_1^\phi=-\dot{r}S_1^{r\phi}-r^2\dot\theta S_1^{\theta\phi}$ and
$(S_1n_{12}v_1)=r^2\dot\theta S_1^{r\theta}+r^2\dot\phi\sin^2\theta
S_1^{r\phi}$. Plugging this into Eq.~\eqref{PNtestmasslimitnospin} leads to
the explicit expressions
\begin{subequations}\label{Schwarschild}
\begin{align}
\label{Schwarschildrpp}
\nonumber (\ddot{r})_S&= \left[
  3 \frac{G}{c^3 r}\frac{m_2}{m_1} -\frac{3}{2}\frac{G}{c^5 r}\frac{m_2}{m_1}
  \left(3\dot{r}^2+r^2\dot\theta^2+r^2\dot\phi^2\sin^2\theta\right)-
6\frac{G^2}{c^5 r^2}\frac{m_2^2}{m_1}\right.\\
\nonumber & +\frac{G}{c^7 r}\frac{m_2}{m_1} \left(
  \frac{9}{8}\dot{r}^4+\frac{3}{4}r^2\dot{r}^2\dot\theta^2-
\frac{3}{8}r^4\dot\theta^4+\frac{3}{4}
  r^2\dot{r}^2\dot\phi^2\sin^2\theta-\frac{3}{4}
  r^4\dot\theta^2\dot\phi^2\sin^2\theta-\frac{3}{8} r^4\dot\phi^4\sin^4\theta
\right)\\
&\left.-3\frac{G^2}{c^7 r^2}\frac{m_2^2}{m_1} 
\left(3\dot{r}^2+r^2\dot\theta^2+r^2\dot\phi^2\sin^2\theta\right)+
\frac{15}{2}\frac{G^3}{c^7 r^3}\frac{m_2^3}{m_1}\right]
\left(\dot\theta S_1^{r\theta}+\dot\phi\sin^2\theta S_1^{r\phi}\right)\;,\\
\label{Schwarschildthetapp}
\nonumber (\ddot{\theta})_S&= 
-3\frac{G}{c^3 r}\frac{m_2}{m_1} \dot\phi 
\sin^2\theta S_1^{\theta\phi}-3\frac{G}{c^5 r}\frac{m_2}{m_1}
\left(\dot{r}\dot\theta^2 S_1^{r\theta}+
\dot{r}\dot{\theta}\dot{\phi}\sin^2\theta S_1^{r\phi}-
\frac{1}{2}v^2\dot\phi\sin^2\theta S_1^{\theta\phi}\right)\\ 
\nonumber & +6\frac{G^2}{c^5 r^2}
\frac{m_2^2}{m_1}\dot\phi\sin^2\theta S_1^{\theta\phi} +
\frac{3}{2}\frac{G}{c^7 r}\frac{m_2}{m_1} v^2
\left(\dot{r}\dot\theta^2 S_1^{r\theta}+
\dot{r}\dot{\theta}\dot{\phi}\sin^2\theta S_1^{r\phi}+
\frac{1}{4}v^2\dot\phi\sin^2\theta S_1^{\theta\phi}\right)\\
&-6\frac{G^2}{c^7 r^2}\frac{m_2^2}{m_1} 
\left(\dot{r}\dot\theta^2 S_1^{r\theta}+
\dot{r}\dot{\theta}\dot{\phi}\sin^2\theta S_1^{r\phi}-
\frac{1}{2}v^2\dot\phi\sin^2\theta S_1^{\theta\phi}\right)-
\frac{15}{2}\frac{G^3}{c^7 r^3}\frac{m_2^3}{m_1} 
\dot\phi\sin^2\theta S_1^{\theta\phi}\;,\\
\label{Schwarschildphipp}
\nonumber (\ddot{\phi})_S&= 
3\frac{G}{c^3 r}\frac{m_2}{m_1} \dot\theta S_1^{\theta\phi}-
3\frac{G}{c^5 r}\frac{m_2}{m_1}\left(\dot{r}\dot\theta\dot\phi S_1^{r\theta}+
\dot{r}\dot{\phi}^2\sin^2\theta S_1^{r\phi}+
\frac{1}{2}\dot\theta  v^2 S_1^{\theta\phi}\right)\\ 
\nonumber & -6\frac{G^2}{c^5 r^2}\frac{m_2^2}{m_1}\dot\theta S_1^{\theta\phi}+
\frac{3}{2}\frac{G}{c^7 r}\frac{m_2}{m_1}v^2 
\left(\dot{r}\dot\theta\dot\phi S_1^{r\theta}+
\dot{r}\dot\phi^2 \sin^2\theta S_1^{r\phi}-
\frac{1}{4}v^2 \dot\theta S_1^{\theta\phi}\right)\\
&-6\frac{G^2}{c^7 r^2}\frac{m_2^2}{m_1} 
\left(\dot{r}\dot\theta\dot\phi S_1^{r\theta}+
\dot{r}\dot{\phi}^2\sin^2\theta S_1^{r\phi}+
\frac{1}{2}v^2\dot\theta S_1^{\theta\phi}\right)+
\frac{15}{2}\frac{G^3}{c^7 r^3}\frac{m_2^3}{m_1} \dot\theta S_1^{\theta\phi}\;,
\end{align}
\end{subequations}
where we have used the shorthand notation $v^2\equiv
(\dot{r}^2+r^2\dot\theta^2+r^2\dot\phi^2\sin^2\theta)$. 

We want now to compare these results to their counterparts as given by the
Papapetrou equation \eqref{papeomlinear} 
written in the Schwarzschild background and in harmonic coordinates. A set of
such coordinates and the corresponding form of the metric can be readily
obtained by setting $a=0$ in Eqs.~\eqref{Kerrharmcoord} and
\eqref{linelement}. We use \eqref{eomcoordinate}, including this time the
Papapetrou force $\calF^{\mu}$, and we obtain expressions for $\ddot{r}$,
$\ddot{\theta}$ and $\ddot{\phi}$ which, when expanded at the 3.5PN order, take
exactly the form of Eqs.~\eqref{Schwarschild}.

\subsection{Equivalence to ADM results}
\label{subsecADM}

In this section, we compare our results to the ones obtained in
Ref.~\cite{HS11spinorbit} by a completely different method, using a reduced
Hamiltonian formalism in ADM-type coordinates. In the following, we will
denote all ADM variables with an overline. We follow the method of
\cite{DJSspin}, which compared to the harmonic-coordinate results in Paper~I
at the next-to-leading order $\calO(5)$. In this formalism, the canonical
structure is defined for the variables $\ov{\bfx}_{a}$ (positions of the two
bodies), $\ov{\bfp}_{a}$ (canonical momenta) and $\overline{\bfS}_{a}$
(canonical spins). Notice that these canonical spins are of conserved norm.
The Poisson brackets for these variables read:
\begin{subequations}
\begin{align}
	\Bigl\{ \ov{x}_{a}^{i}, \ov{p}_{b}^{j}\Bigr\} &= \delta_{ab}\delta^{ij}
    \;,\\ \Bigl\{ \ov{S}_{a}^{i}, \ov{S}_{b}^{j}\Bigr\} &= c \,\delta_{ab}
    \varepsilon^{ijk}\ov{S}^{k} \;,
\end{align}
\end{subequations}
and all the other Poisson brackets vanish. Beware of the additional $c$ factor
appearing in the second bracket: our PN counting is different from that in
\cite{HS11spinorbit}, and we have set $\ov{S}^{i}=c \hat{S}_{\text{HS}}^{i}$
to keep close to the convention we use, see Eq.~\eqref{spinPNcounting}. One
may as well employ a conserved-norm antisymmetric spin tensor defined
(exactly) by:
\begin{equation}\label{defSijcanonical}
	\ov{S}^{ij} \equiv \varepsilon^{ijk}\ov{S}^{k} \;.
\end{equation}

The dynamics is entirely contained in the Hamiltonian of the two-body system,
expressed in terms of these canonical variables. Besides the recent result for
the spin-orbit 3.5PN spin-orbit contributions \cite{HS11spinorbit}, the
non-spin part of the Hamiltonian up to 2PN order can be found for instance in
\cite{DJSequiv}, and the spin-orbit part up to 2.5PN order is given in
\cite{DJSspin}. As explained before, since we are working at 2PN relative
order with respect to the leading order spin-orbit contribution, we will need
to know the non-spin part of the dynamics only up to 2PN order. The time
derivative of any quantity $f$ is obtained as:
\begin{equation}
	\frac{\ud f}{\ud t} = \left\{ f,H \right\} + 
    \frac{\partial f}{\partial t} \;,
\end{equation}
where the second term accounts for a possible explicit time dependence of $f$,
absent in our problem. Beware again that, because of the additional $c$
factor in the Poisson brackets for the spin, the PN truncation of the
Hamiltonian must be handled carefully.

It is well known that, when comparing the results of the two formalisms, it is
necessary to perform a contact transformation of the worldlines of the
particles, \text{i.e.} a time-dependent shift of the worldlines that is
not a global coordinate transformation. Thus, one must keep in mind that the
positions $\bfy_{a}$ and $\ov{\bfx}_{a}$ differ at higher-order. We write this
symbolically as
\begin{equation}\label{ytoadm}
	\bfy = \mathbf{Y}(\ov{\bfx},\ov{\bfp},\ov{\bfS}) \;.
\end{equation}
The link between the harmonic velocity $\bfv$ and canonical momentum
$\ov{\bfp}$ is obtained by:
\begin{equation}\label{vtoadm}
	\bfv = \bfV(\ov{\bfx},\ov{\bfp},\ov{\bfS}) = 
    \left\{ \mathbf{Y}(\ov{\bfx},\ov{\bfp},\ov{\bfS}), H \right\} \;.
\end{equation}
As for the acceleration, we have:
\begin{equation}\label{atoadmdirect}
	\mathbf{a} = \mathbf{A}(\ov{\bfx},\ov{\bfp},\ov{\bfS}) = 
\left\{ \left\{ \mathbf{Y}(\ov{\bfx},\ov{\bfp},\ov{\bfS}), H \right\} ,
H \right\} \;.
\end{equation}
On the other hand, we may use \eqref{ytoadm} and \eqref{vtoadm} and the link
between the spin variables \eqref{S1toADMstruct} to translate the
harmonic-coordinates result \eqref{a1iS7} for the acceleration in terms of the
ADM variables as:
\begin{equation}\label{atoadmindirect}
	\mathbf{a} = \mathbf{a}(\bfy,\bfv,S) = 
\mathbf{a}(\mathbf{Y}(\ov{\bfx},\ov{\bfp},
\ov{\bfS}),\bfV(\ov{\bfx},\ov{\bfp},\ov{\bfS}),
S(\ov{\bfx},\ov{\bfp},\ov{\bfS})) \;,
\end{equation}
and this expression must be identical to \eqref{atoadmdirect}. Thus, one is
led to look for an extension of the contact transformation \eqref{ytoadm}
that realizes this identity. Focusing on the non-spin and spin-orbit parts,
the structure of this contact transformation is the following:
\begin{equation}\label{Y1struct}
	\mathbf{Y}_{1} = \ov{\bfx}_{1} +
\frac{1}{c^3} \mathop{\mathbf{Y}}_{S}{}_{1}^{\!\mathrm{1.5PN}} + 
\frac{1}{c^4} {\mathbf{Y}}_{1}^{\mathrm{2PN}} + 
\frac{1}{c^5} \mathop{\mathbf{Y}}_{S}{}_{1}^{\!\mathrm{2.5PN}} +  
\frac{1}{c^6} {\mathbf{Y}}_{1}^{\mathrm{3PN}} + 
\frac{1}{c^7} \mathop{\mathbf{Y}}_{S}{}_{1}^{\!\mathrm{3.5PN}} + 
\mathcal{O}\left(\frac{1}{c^8}\right) \,.
\end{equation}

The lower-order already known expressions for this contact transformation can
be found in \cite{DJSequiv} and \cite{DJSspin}. Notice that, as when using the
Hamiltonian to compute time derivatives in the ADM setting, we need the
non-spin part of this contact transformation up to 2PN only. To present these
formulae, we use the canonical spin tensor variable \eqref{defSijcanonical} to
get the same index structure as in the other results of the present paper, and
we adopt the convenient notation $\ov{\pi}_{a} \equiv \ov{p}_{a}/m_{a}$ to
shorten the expressions. We find:
\begin{subequations}
\begin{align}\label{}
	{\mathbf{Y}}_{1}^{\mathrm{2PN}} &= G m_{2} 
\left[ \frac{1}{2} \ov{\pi}_{1}^{i}(\ov{n}_{12}\ov{\pi}_{2}) -
\frac{7}{4} \ov{\pi}_{2}^{i} (\ov{n}_{12}\ov{\pi}_{2}) + 
\frac{1}{8} \ov{n}_{12}^{i} \left( 5\ov{\pi}_{2}^{2} - 
(\ov{n}_{12}\ov{\pi}_{2})^{2} \right) \right] \nonumber \\
	& \quad + \frac{G^{2}m_{2}}{4\ov{r}_{12}} \ov{n}_{12}^{i} 
\left( 7m_{1} + m_{2} \right) \;,\\
        \label{}
	m_{1} \mathop{\mathbf{Y}}_{S}{}_{1}^{\!\mathrm{1.5PN}} &= - 
\frac{1}{2}\ov{S}_{1}^{ij}\ov{\pi}_{1}^{j} \;,\\
        \label{}
	m_{1} \mathop{\mathbf{Y}}_{S}{}_{1}^{\!\mathrm{2.5PN}} &= 
\frac{\ov{\pi}_{1}^{2}}{8} \ov{S}_{1}^{ij}\ov{\pi}_{1}^{j} \nonumber \\
	& + \frac{G}{\ov{r}_{12}} \left[ m_{2}\ov{S}_{1}^{ij}\ov{\pi}_{1}^{j} + 
m_{1} \left( -\frac{3}{2}\ov{S}_{2}^{ij}\ov{\pi}_{2}^{j} + 
(\ov{n}_{12}\ov{\pi}_{2}) \ov{S}_{2}^{ij}\ov{n}_{12}^{j} + 
\frac{1}{2} (\ov{S}_{2}\ov{n}_{12}\ov{\pi}_{2})\ov{n}_{12}^{i}  \right)\right] \;.
\end{align}
\end{subequations}

We need also the conversion rule between the harmonic spin tensor and the ADM
canonical spin, say $S=S(\ov{\bfx},\ov{\bfp},\ov{\bfS})$.
The conversion rule from the conserved norm
variable $\bfS^{c}_{\text{FBB}}$ to $\ov{\bfS}$ is given in Eqs.~(6.4) and
(6.5) of \cite{DJSspin}, and the conversion rule between this conserved norm
spin and $\bfS_{\text{FBB}}$ is given in Eq.~(7.4) of \cite{BBF06}. Finally,
knowing the link between $\bfS_{\text{FBB}}$ and our spin tensor, 
Eq.~\eqref{S1iFBBtoS1ijstruct} in Appendix~\ref{appSiSij}, we
get:
\begin{equation}\label{S1toADMstruct}
	S_{1}^{ij} = \ov{S}_{1}^{ij} + 
\frac{1}{c^2} \mathop{\ov{\Sigma}}_{\!\mathrm{1PN}}{}_{1}^{ij} + 
\frac{1}{c^4} \mathop{\ov{\Sigma}}_{\!\mathrm{2PN}}{}_{1}^{ij} + 
\mathcal{O}\left( 5 \right) \;,
\end{equation}
where
\begin{subequations}\label{S1toADMstructexpl}
\begin{align}\label{S1toADM2}
	\mathop{\ov{\Sigma}}_{\!\mathrm{1PN}}{}_{1}^{ij} =& -
\ov{\pi}_{1}^{[i}\ov{S}_{1}^{j]k}\ov{\pi}_{1}^{k} -  
\frac{2 G m_{2}}{\ov{r}_{12}} \ov{S}_{1}^{ij} \;,\\
\label{S1toADM4}
	\mathop{\ov{\Sigma}}_{\!\mathrm{2PN}}{}_{1}^{ij} =&  
\Sigma^{ij}_{0,0} + \frac{G}{\ov{r}_{12}} \Sigma^{ij}_{0,1} m_{2} +
\frac{G^{2}}{\ov{r}_{12}^{2}} \left( \Sigma^{ij}_{0,2} m_{2}^{2} + 
\Sigma^{ij}_{1,1} m_{1}m_{2} \right)  \;,\\
    \Sigma^{ij}_{0,0} =& 
\frac{1}{4} \ov{\pi}_{1}^{2} \ov{\pi}_{1}^{[i}\ov{S}_{1}^{j]k}\ov{\pi}_{1}^{k} \;,\\
    \Sigma^{ij}_{0,1} =& - 
(\ov{n}_{12}\ov{\pi}_{2}) \ov{n}_{12}^{[i}\ov{S}_{1}^{j]k}\ov{\pi}_{1}^{k} + 
2 (\ov{n}_{12}\ov{\pi}_{2}) \ov{n}_{12}^{[i}\ov{S}_{1}^{j]k}\ov{\pi}_{2}^{k} + 
4 \ov{\pi}_{1}^{[i}\ov{S}_{1}^{j]k}\ov{\pi}_{1}^{k} - 
2 (\ov{n}_{12}\ov{\pi}_{2}) \ov{\pi}_{2}^{[i}\ov{S}_{1}^{j]k}\ov{n}_{12}^{k} 
\nonumber \\
& - 7 \ov{\pi}_{2}^{[i}\ov{S}_{1}^{j]k}\ov{\pi}_{1}^{k} + 
4 \ov{\pi}_{2}^{[i}\ov{S}_{1}^{j]k}\ov{\pi}_{2}^{k} + 
(\ov{n}_{12}\ov{\pi}_{2})^2 \ov{S}_{1}^{ij} \;,\\
    \Sigma^{ij}_{0,2} =& \ov{n}_{12}^{[i}\ov{S}_{1}^{j]k}\ov{n}_{12}^{k} + 
3 \ov{S}_{1}^{ij} \;,\\
    \Sigma^{ij}_{1,1} =& -8 \ov{n}_{12}^{[i}\ov{S}_{1}^{j]k}\ov{n}_{12}^{k} - 
\ov{S}_{1}^{ij} \;.
\end{align}
\end{subequations}

Using the same structural hypothesis as in \cite{DJSspin} (namely that in
$m_{1}\mathbf{Y}_{1}$ terms such as $m_{1}^{n}S_{1}$ and $m_{2}^{n}S_{2}$ are
forbidden, where $n$ is the total power of mass in the term), we build a
putative spin contribution to the contact transformation with 112 unknown
coefficients. Then, requiring the identity of the expressions
\eqref{atoadmdirect} and \eqref{atoadmindirect} above, we find a unique
solution for these coefficients, among which 69 remain in the final result,
which fixes uniquely the contact transformation. Thus, we conclude that our
result and the one 
obtained in the ADM formalism in Ref.~\cite{HS11spinorbit} are
equivalent.\footnote{We found a typographical error in the originally
  published version of 
  the Hamiltonian (5) in Ref.~\cite{HS11spinorbit}: the coefficient in front
  of the term
  $\frac{G}{r_{12}^{2}}
  \frac{(\mathbf{n}_{12}\mathbf{P}_{1})\mathbf{P}_{2}^{2}}{m_{1}^{2}m_{2}^{2}}
  ((\mathbf{P}_{1}\times
  \mathbf{P}_{2})\hat{\mathbf{S}}_{1})$ should read $-\frac{5}{16}$ instead of
  $-\frac{15}{16}$.}

We now give explicitly the 3.5PN spin-orbit extension of the contact
transformation which reads:
\begin{equation}\label{Y1iS7}
	m_{1}\mathop{Y^i}_{S}{}_{\!\mathrm{3.5PN}} = -
\frac{\ov{\pi}_{1}^{4}}{16} \ov{S}_{1}^{ij}\ov{\pi}_{1}^{j} + 
\frac{G}{r_{12}}\left[ \lambda^{i}_{0,1} m_{2} + 
\lambda^{i}_{1,0} m_{1} \right] + 
\frac{G^{2}}{r_{12}^{2}} \left[ \lambda^{i}_{0,2}m_{2}^{2} + 
\lambda^{i}_{1,1} m_{1}m_{2} + \lambda^{i}_{2,0} m_{1}^{2} \right] \; ,
\end{equation}
where
\begin{subequations}
\begin{align}
    \lambda^{i}_{0,1} &= \ov{S}_{1}^{ij}\ov{n}_{12}^{j} 
\left[ \frac{3}{16} (\ov{n}_{12}\ov{\pi}_{1}) (\ov{n}_{12}\ov{\pi}_{2})^2 - 
\frac{3}{8} (\ov{n}_{12}\ov{\pi}_{2})^3 - 
\frac{1}{4} (\ov{n}_{12}\ov{\pi}_{2}) \ov{\pi}_{1}^{2} + 
\frac{1}{4} (\ov{n}_{12}\ov{\pi}_{1}) (\ov{\pi}_{1}\ov{\pi}_{2}) \right.
\nonumber \\
& \qquad \qquad \quad \left. + 
\frac{11}{8} (\ov{n}_{12}\ov{\pi}_{2}) (\ov{\pi}_{1}\ov{\pi}_{2}) - 
\frac{21}{16} (\ov{n}_{12}\ov{\pi}_{1}) \ov{\pi}_{2}^{2} - 
\frac{1}{8} (\ov{n}_{12}\ov{\pi}_{2}) \ov{\pi}_{2}^{2} \right] \nonumber \\
& + \ov{S}_{1}^{ij}\ov{\pi}_{1}^{j} 
\left[ \frac{1}{4} (\ov{n}_{12}\ov{\pi}_{1}) (\ov{n}_{12}\ov{\pi}_{2}) - 
\frac{25}{16} (\ov{n}_{12}\ov{\pi}_{2})^2 - 
\frac{1}{2} \ov{\pi}_{1}^{2} - \frac{1}{4} (\ov{\pi}_{1}\ov{\pi}_{2}) + 
\frac{21}{16} \ov{\pi}_{2}^{2} \right] \nonumber \\
& + \ov{S}_{1}^{ij}\ov{\pi}_{2}^{j} 
\left[ -\frac{1}{4} (\ov{n}_{12}\ov{\pi}_{1})^2 + 
\frac{11}{8} (\ov{n}_{12}\ov{\pi}_{1}) (\ov{n}_{12}\ov{\pi}_{2}) + 
\frac{5}{8} (\ov{n}_{12}\ov{\pi}_{2})^2 + \frac{1}{4} \ov{\pi}_{1}^{2} - 
\frac{7}{8} (\ov{\pi}_{1}\ov{\pi}_{2}) - 
\frac{1}{8} \ov{\pi}_{2}^{2} \right] \nonumber \\
& + (\ov{S}_{1}\ov{n}_{12}\ov{\pi}_{1})\ov{n}_{12}^{i} 
\left[ \frac{1}{4} (\ov{\pi}_{1}\ov{\pi}_{2}) -  
\ov{\pi}_{2}^{2} \right] + 
\frac{1}{2} (\ov{n}_{12}\ov{\pi}_{2})
(\ov{S}_{1}\ov{n}_{12}\ov{\pi}_{1})\ov{\pi}_{2}^{i} \nonumber \\
& + (\ov{S}_{1}\ov{n}_{12}\ov{\pi}_{2})\ov{n}_{12}^{i} 
\left[ -\frac{3}{4} (\ov{n}_{12}\ov{\pi}_{2})^2 - 
\frac{1}{4} \ov{\pi}_{1}^{2} + (\ov{\pi}_{1}\ov{\pi}_{2}) + 
\frac{1}{4} \ov{\pi}_{2}^{2} \right] + 
\frac{1}{4} (\ov{n}_{12}\ov{\pi}_{1})
(\ov{S}_{1}\ov{n}_{12}\ov{\pi}_{2})\ov{\pi}_{1}^{i} \nonumber \\ 
& + (\ov{S}_{1}\ov{n}_{12}\ov{\pi}_{2})\ov{\pi}_{2}^{i} 
\left[ - (\ov{n}_{12}\ov{\pi}_{1}) - 
\frac{1}{2} (\ov{n}_{12}\ov{\pi}_{2}) \right] + 
(\ov{S}_{1}\ov{\pi}_{1}\ov{\pi}_{2})\ov{n}_{12}^{i} 
\left[ \frac{1}{4} (\ov{n}_{12}\ov{\pi}_{1}) - 
\frac{3}{2} (\ov{n}_{12}\ov{\pi}_{2}) \right] \; ,\\
    \lambda^{i}_{1,0} &= \ov{S}_{2}^{ij}\ov{n}_{12}^{j} 
\left[ -\frac{3}{4} (\ov{n}_{12}\ov{\pi}_{1}) (\ov{n}_{12}\ov{\pi}_{2})^2 - 
\frac{3}{4} (\ov{n}_{12}\ov{\pi}_{2})^3 - 
\frac{5}{8} (\ov{n}_{12}\ov{\pi}_{2}) (\ov{\pi}_{1}\ov{\pi}_{2}) \right.
\nonumber \\ 
& \qquad \qquad \quad \left. + 
\frac{3}{8} (\ov{n}_{12}\ov{\pi}_{1}) \ov{\pi}_{2}^{2} + 
\frac{1}{4} (\ov{n}_{12}\ov{\pi}_{2}) \ov{\pi}_{2}^{2} \right] \nonumber \\
& + \ov{S}_{2}^{ij}\ov{\pi}_{1}^{j} 
\left[ \frac{1}{8} (\ov{n}_{12}\ov{\pi}_{2})^2 - 
\frac{1}{8} \ov{\pi}_{2}^{2} \right] + 
\ov{S}_{2}^{ij}\ov{\pi}_{2}^{j} 
\left[ \frac{1}{4} (\ov{n}_{12}\ov{\pi}_{1}) (\ov{n}_{12}\ov{\pi}_{2}) + 
\frac{7}{16} (\ov{n}_{12}\ov{\pi}_{2})^2 + 
\frac{1}{4} (\ov{\pi}_{1}\ov{\pi}_{2}) + 
\frac{5}{16} \ov{\pi}_{2}^{2} \right] \nonumber \\
& + (\ov{S}_{2}\ov{n}_{12}\ov{\pi}_{1})\ov{n}_{12}^{i} 
\left[ \frac{3}{4} (\ov{n}_{12}\ov{\pi}_{2})^2 - 
\frac{1}{8} \ov{\pi}_{2}^{2} \right] + 
\frac{3}{8} (\ov{n}_{12}\ov{\pi}_{2})
(\ov{S}_{2}\ov{n}_{12}\ov{\pi}_{1})\ov{\pi}_{2}^{i} \nonumber \\ 
& + (\ov{S}_{2}\ov{n}_{12}\ov{\pi}_{2})\ov{n}_{12}^{i} 
\left[ -\frac{9}{8} (\ov{n}_{12}\ov{\pi}_{1}) (\ov{n}_{12}\ov{\pi}_{2}) - 
\frac{9}{16} (\ov{n}_{12}\ov{\pi}_{2})^2 - 
\frac{1}{4} (\ov{\pi}_{1}\ov{\pi}_{2}) - 
\frac{1}{16} \ov{\pi}_{2}^{2} \right] \nonumber \\
& + \frac{5}{8} (\ov{n}_{12}\ov{\pi}_{2}) 
(\ov{S}_{2}\ov{n}_{12}\ov{\pi}_{2})\ov{\pi}_{1}^{i} - 
\frac{1}{8} (\ov{n}_{12}\ov{\pi}_{2})
(\ov{S}_{2}\ov{n}_{12}\ov{\pi}_{2})\ov{\pi}_{2}^{i} +
 \frac{1}{2} (\ov{n}_{12}\ov{\pi}_{2})
 (\ov{S}_{2}\ov{\pi}_{1}\ov{\pi}_{2})\ov{n}_{12}^{i} \; ,\\
    \lambda^{i}_{0,2} &= -
\frac{1}{2} (\ov{n}_{12}\ov{\pi}_{2}) \ov{S}_{1}^{ij}\ov{n}_{12}^{j} - 
\frac{11}{8} \ov{S}_{1}^{ij}\ov{\pi}_{1}^{j} + 
\frac{1}{4} (\ov{S}_{1}\ov{n}_{12}\ov{\pi}_{1})\ov{n}_{12}^{i} \; ,\\
    \lambda^{i}_{1,1} &= \ov{S}_{1}^{ij}\ov{n}_{12}^{j} 
\left[ -\frac{45}{8} (\ov{n}_{12}\ov{\pi}_{1}) + 
\frac{43}{8} (\ov{n}_{12}\ov{\pi}_{2}) \right] + 
\frac{35}{8} \ov{S}_{1}^{ij}\ov{\pi}_{1}^{j} - 
\frac{139}{16} \ov{S}_{1}^{ij}\ov{\pi}_{2}^{j} - 
\frac{73}{8} (\ov{S}_{1}\ov{n}_{12}\ov{\pi}_{1})\ov{n}_{12}^{i} \nonumber \\
& + \frac{23}{2} (\ov{S}_{1}\ov{n}_{12}\ov{\pi}_{2})\ov{n}_{12}^{i} -
4 (\ov{n}_{12}\ov{\pi}_{2}) \ov{S}_{2}^{ij}\ov{n}_{12}^{j} + 
\frac{49}{16} \ov{S}_{2}^{ij}\ov{\pi}_{2}^{j} - 
\frac{15}{8} (\ov{S}_{2}\ov{n}_{12}\ov{\pi}_{2})\ov{n}_{12}^{i} \; ,\\
    \lambda^{i}_{2,0} &= \ov{S}_{2}^{ij}\ov{n}_{12}^{j} 
\left[ -\frac{39}{8} (\ov{n}_{12}\ov{\pi}_{1}) + 
\frac{25}{8} (\ov{n}_{12}\ov{\pi}_{2}) \right] + 
3 \ov{S}_{2}^{ij}\ov{\pi}_{1}^{j} - 
\frac{1}{4} \ov{S}_{2}^{ij}\ov{\pi}_{2}^{j} - 
\frac{57}{8} (\ov{S}_{2}\ov{n}_{12}\ov{\pi}_{1})\ov{n}_{12}^{i} \nonumber \\
& + \frac{23}{8} (\ov{S}_{2}\ov{n}_{12}\ov{\pi}_{2})\ov{n}_{12}^{i}\; .
\end{align}
\end{subequations}
 %
 

\section{Conclusion}
\label{Conclusion}

In this work, we computed the next-to-next-to-leading order spin-orbit
contributions to the equations of motion of compact binaries. Those are of
3.5PN order for maximally spinning objects, thus improving our knowledge of
the dynamics of such systems at this order. Our result was tested by checking
the existence of a conserved energy, the manifest Lorentz invariance of the
equations of motion, and the agreement of the test-mass limit with the motion
of a spinless test particle around a Kerr black hole as well as that of a
spinning test particle around a Schwarzschild black hole. We also recover and
confirm the result obtained previously in Ref.~\cite{HS11spinorbit} using a
reduced Hamiltonian method in ADM-type coordinates, extending the contact
transformation that makes the link between the two formalisms.

We leave for future work the study of the 3PN precession equation for the
spins, the construction of the conserved quantities other than the energy
(total angular momentum, linear momentum, center-of-mass integral), the
center-of-mass reduction of the equations of motion as well as the further
reduction of the dynamics to quasi-circular orbits. We shall also provide the
spin-orbit contributions to the components of the near-zone metric itself and
its value at the location of the particles. Most importantly, this work opens
the way to the computation of the 3.5PN spin-orbit contributions to the energy
flux emitted by the binary in the form of gravitational waves, and of the
corresponding contribution in the phase of the GW signal, which should improve
the templates used by current and future GW detectors.


\appendix

\section{Link between different spin variables} 
\label{appSiSij}

In this Appendix we provide the explicit conversion rule from the spin
variable used in Paper~I, which we denote $S_{\text{FBB}}^{i}$, to our spin
tensor variable $S^{ij}$. This rule allows one to readily translate the
lower-order results of Paper~I in terms of our variables for comparison. We
adopt the notation $(\varepsilon a S)$ for $\varepsilon^{jkl}a^{j}S^{kl} $,
for any vector $a$, with the indices in this precise order:
\begin{equation}\label{S1iFBBtoS1ijstruct}
\mathop{S}_{\text{FBB}}{}_{1}^{i} = \frac{1}{2}\varepsilon^{ijk}S_{1}^{jk} + 
\frac{1}{c^2} \mathop{\Sigma}_{\!\mathrm{1PN}}{}_{1}^{i} + 
\frac{1}{c^4} \mathop{\Sigma}_{\!\mathrm{2PN}}{}_{1}^{i}  + 
\mathcal{O}\left( 5 \right) \;,
\end{equation}
where
\begin{subequations}
\begin{align}\label{S1iFBBtoS1ij4}
\mathop{\Sigma}_{\!\mathrm{1PN}}{}_{1}^{i} &= -
\frac{1}{4}v_{1}^{2}\varepsilon^{ijk}S_{1}^{jk} + 
\frac{1}{2} (\varepsilon v_{1} S_{1}) v_{1}^{i} + 
\frac{G m_{2}}{2 r_{12}} \varepsilon^{ijk}S_{1}^{jk} \;,\\
\mathop{\Sigma}_{\!\mathrm{2PN}}{}_{1}^{i} &= -
\frac{1}{16}v_{1}^{4}\varepsilon^{ijk}S_{1}^{jk} + 
\frac{1}{4} (\varepsilon v_{1} S_{1})v_{1}^{2} v_{1}^{i} + 
\frac{G m_{2}}{r_{12}} \left[ \varepsilon^{ijk}S_{1}^{jk}
\left( -\frac{5}{4}v_{1}^{2} + v_{2}^{2} - 
\frac{1}{4}(n_{12}v_{2})^{2} \right) \right. \nonumber \\ 
	& \left. + \frac{5}{2} (\varepsilon v_{1} S_{1}) v_{1}^{i} - 
2 (\varepsilon v_{1} S_{1}) v_{2}^{i} - 
2 (\varepsilon v_{2} S_{1}) v_{2}^{i} \right] + 
\frac{G^{2}m_{1}m_{2}}{r_{12}^{2}} \left[
  -\frac{7}{4}\varepsilon^{ijk}S_{1}^{jk} + 
4(\varepsilon n_{12} S_{1})n_{12}^{i}  \right] \nonumber \\
	& + \frac{G^{2}m_{2}^{2}}{r_{12}^{2}} 
\left[ \frac{1}{4}\varepsilon^{ijk}S_{1}^{jk} - 
\frac{1}{2}(\varepsilon n_{12} S_{1})n_{12}^{i}  \right] \;.
\end{align}
\end{subequations}
On the other hand the link between our spin tensor and the spin variable used
in the ADM-Hamiltonian work~\cite{HS11spinorbit} is provided in
Eqs.~\eqref{S1toADMstruct}--\eqref{S1toADMstructexpl}.


\bibliography{ListeRef}

\end{document}